\documentclass[aps,pra,notitlepage,superscriptaddress,showpacs,nofootinbib]{revtex4-1}
\usepackage{amsmath}
\usepackage{amssymb}
\usepackage{amsfonts}
\usepackage{enumerate}
\usepackage{mathrsfs}
\usepackage{colortbl}
\usepackage{array}
\usepackage{color}
\usepackage{indentfirst}
\usepackage{amssymb}
\usepackage{amsmath}
\usepackage{tikz}
\usepackage{amscd}
\usepackage{amsthm}
\usepackage{graphicx}
\usepackage{mathdots}
\usepackage{epstopdf}
\usepackage{enumerate}
\usepackage{changepage}

\usepackage[
colorlinks,
linkcolor = blue,
citecolor = blue,
urlcolor = blue]{hyperref}
\def \qed {\hfill \vrule height7pt width 7pt depth 0pt}

\newtheorem{theorem}{Theorem}
\newtheorem{corollary}{Corollary}
\newtheorem{lemma}{Lemma}
\newtheorem{example}{Example}
\newtheorem{proposition}{Proposition}

\usepackage{dcolumn}
\usepackage{bm}

\usepackage{multirow}
\usepackage{colortbl}
\definecolor{tabcolor}{rgb}{0.245,0.245,0.245}
\usepackage{lipsum}
\definecolor{mygray}{gray}{.9}
\definecolor{mypink}{rgb}{.99,.91,.95}
\definecolor{mycyan}{cmyk}{.3,0,0,0}

\begin{document}

\title{Planar $k$-Uniform States: a Generalization of  Planar Maximally Entangled States}
\author{Yan-Ling Wang }\email{wangylmath@yahoo.com}

\affiliation{School of Computer Science and Techonology, Dongguan University of Technology, Dongguan 523808, China}

\begin{abstract}
Recently,   Doroudiani  and  Karimipour [\href{https://journals.aps.org/pra/abstract/10.1103/PhysRevA.102.012427}{Phys. Rev. A \textbf{102} 012427(2020)}] proposed the notation of  planar maximally entangled (PME) states which are a wider class of multipartite entangled states than absolutely maximally entangled (AME) states. There they presented their constructions in the multipartite systems but the  number of particles is restricted to be even. Here we  first solve the remaining cases, i.e., constructions of  planar maximally entangled states on systems with odd number of particles. In addition,  we generalized the PME to the planar $k$-uniform states whose   reductions to any adjacent $k$ parties along a circle  of $N$ parties are maximally mixed.  We presented a method to construct  sets of   planar $k$-uniform states which have minimal support.  
\end{abstract}

 \maketitle

\section{Introduction}
\setlength{\parindent}{1em}
Since the surprising  application named quantum teleportation \cite{Bennett93,Bouwmeester97} of quantum entanglement  has been found, lots of effort have been made to understanding and quantifying the entanglement for bipartite entangled states \cite{Horodecki09}.  Despite the fruitful knowledge  on   pure bipartite entangled states,  we know little about the entanglement among  multipartite systems. However, multipartite entanglement is also an important resource   in many  application scenarios such as quantum networks \cite{peres},   quantum metrology \cite{maccone1, metro, noise}, and distributed quantum computing \cite{rausendorf}.  So a well understanding of the structure of the  multipartite  entanglement is an important problem in quantum information theory.

For  bipartite states,   the maximally entangled states are the most significant states among  the set of entangled states.  For
three qubits, the most famous states are GHZ state  and
W state \cite{GHZ90,ciracGHZ}.  These states share a common feature: each local states of them are maximally mix. An interesting problem is to find their analogues for $N$ parties  system. In fact, there has been a  striking definition of  the
``canonical maximally  entangled states which is called absolutely maximally entangled state (AME) \cite{Facchi08}.  Such  state is  maximally  entangled on  every  bipartition of these $N$ subsystems. Equivalently, all its reductions to   $\lfloor \frac{N}{2}\rfloor$ parties are maximally mixed. However,  only a few AME states exist for qubits: 2, 3, 5, 6 qubits and not exist for other cases\cite{Facchi08,raissi17,Raissi17K,Helwig12,Helwig13,Huber17,Huber18}.  Therefore, the  AME states are considered to be a rather limited concept.
 A much more general concept is the $k$-uniform states whose reductions to any $k$ parties are maximally mixed.  There are more techniques and fruitful results on  the construction of  $k$-uniform states. Goyeneche \emph{et al.} \cite{Goyeneche14,Goyeneche15,Goyeneche18} related its construction with  some combinatoric objects such as orthogonal arrays and quantum Latin squares. Recently, more results and some other methods are found for constructing  the $k$-uniform states or even the AME \cite{,Li19,Pang19,Raissi20,Shi20mix,Shi20masking}.

  Quite recently, as the limitation of  the AME, Doroudiani  and  Karimipour \cite{Doroudiani20} proposed a rather general notation   than absolutely maximally entangled (AME) states named planar maximally entangled (PME) states. These states are assumed to be  maximally  entangled on   every   bipartition of  the adjacent subsystems whose underlying topology is a circle. They presented a construction of PME for those systems with  even number of subsystems and pointed out their applications on the teleportation and quantum secret sharing \cite{Hillery99,Cleve99}. Therefore, it is natural to ask whether there exists some PME when the parties number is odd. Moreover, what can we say if we consider the planar $k$-uniform state (state whose  reduction  on    every $k$ ``adjacent" subsystems is maximally mix)? In this paper, we will consider these two problems.

  The remaining of this article is organized as follows.   In Sec. \ref{second}, we give some notation and basic result on bipartite maximally entangled states.   In  Sec. \ref{third}, we will present a construction of the PME for systems with odd number of parties.   Then we generalize the concept PME to planar $k$-uniform state and present a general construction of  planar $k$-uniform state  for every possible case in Sec. \ref{fourth}.  Moreover, we also study some properties of the planar $k$-uniform states. Finally, we conclude   in Sec. \ref{fifth}.

\section{Some preliminary definition, notation and  fact}\label{second}

\begin{figure}[h]
	\centering
	\includegraphics[scale=0.5]{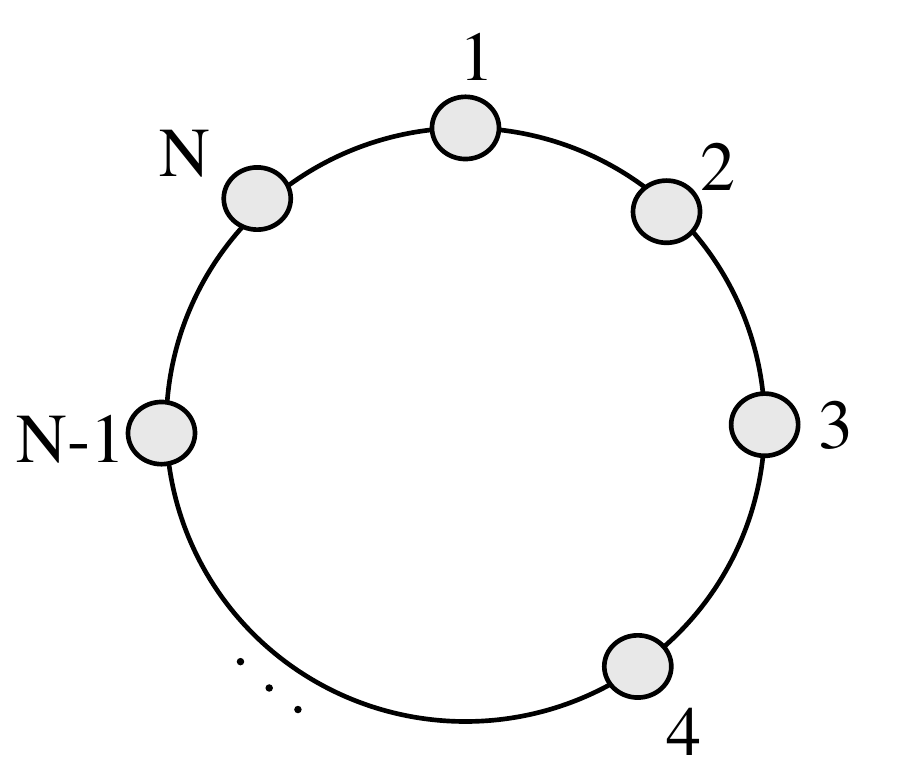} \ \ \ \ \ \ \ \ \
	\caption{ The topology of the quantum particles  throughout this paper. }\label{TP}
\end{figure}

Let $\mathbb{Z}_d$ denote the cyclic  group which  contains $d$ elements, i.e.,  the set $\{0,1,\cdots, d-1\}$ with $i\oplus j \equiv i+j \mod d$.  Throughout this paper, we will consider the $N$ particles quantum system whose underlying topology is a circle (see Fig. \ref{TP}).  That is, each subsystem has two neighbourhood along this circle. We call \emph{Planar Maximally Entangled} (PME) states have the property that any collection of $ {\lfloor\frac{N}{2}\rfloor}$ {\it {adjacent}} particles are in a completely mixed state and by adjacent here we also imply the underlying graph is a circle.

To make  the reductions of some subsystems  to be completely mixed, the following fact is very useful.
 Consider a partition of the particles into parts $A$ and $B$ with $|A|\leq |B|$ and expand a state $|\Psi\rangle$ as
\begin{equation}
\label{basicPsi}
|\Psi\rangle=\sum_{{\bf i}}|{\bf i}\rangle_A|\phi_{\bf i}\rangle_B.
\end{equation}
Then the density matrix of part $A$ is given by
\begin{equation}\label{basicRho}
\rho_A=\sum_{{\bf i},{\bf i'}}|{\bf i}\rangle_A\langle{\bf i'} |\langle \phi_{\bf i}|\phi_{\bf i'}\rangle_B.
\end{equation}

\noindent\textbf{Fact:} For this density matrix to be maximally mixed, it is sufficient to require that the states $\{|{\bf i}\rangle\}$ form a basis for the Hilbert space of part $A$ and the states $\{|\phi_{\bf i}\rangle\}$ are orthogonal and with equal norm.

\section{PMES for systems with odd number of parties }\label{third}

 Before we present our construction of PMEs for systems with odd number of parties we give a brief review of the results and constructions given by Doroudiani and Karimipour in Ref. \cite{Doroudiani20}. They gave a complete characterization of PMEs of four qubits system. In fact, they   showed that there are two families of such states, a two-parameter family and a four-parameter family. After that thet constructed a large family of PMEs  any systems of even number of particles with the same dimension.    Such constructions are inspired by the classical maximally entangled states $|\phi_+\rangle:=\frac{1}{\sqrt{d}}\displaystyle\sum_{i\in\mathbb{Z}_d} |ii\rangle$. As   example, they showed   
\begin{equation}\label{PMEeven}|\Phi_e\rangle:=\prod_{k=1}^n |\phi_+\rangle_{k,n+k} =\frac{1}{d^{n/2}}\sum_{i_1,\cdots,i_n=0}^{{d-1}}(\otimes_{k=1}^n|i_k\rangle_k)\otimes (\otimes_{k=1}^n|i_k\rangle_{n+k})
	\end{equation} 
is always a PME of $2n$ parties of level $d$. Moreover, they found that any PME state of $2n$ parties can be used for quantum state sharing
 between $2n-1$ players so that any adjacent players of size greater than $n$ can recover the state. Given a  PME state $|\Psi\rangle$ of $2n$ parties and suppose `$a$' is one of the $2n$ parties $A=\{a_1,a_2,\cdots,a_{2n}\}$. The coding process can be determined by 
 $$|k\rangle_a\rightarrow |S_k\rangle_{A\setminus \{a\}}, \ \ k=0,1,\cdots,d-1$$
 where $|\Psi\rangle=\frac{1}{\sqrt{d}}\sum_{k=0}^{d-1} |k\rangle_a |S_k\rangle_{A\setminus\{a\}}$.

In the following, we first present two examples of PMEs in the odd number of particles. With these examples in mind, they helps us to understand the general ones. Followed by the fact in Sec. \ref{second} and the corresponding key point below, the constructing states are indeed PMEs.

\begin{figure}[h]
	\centering
	\includegraphics[scale=0.32]{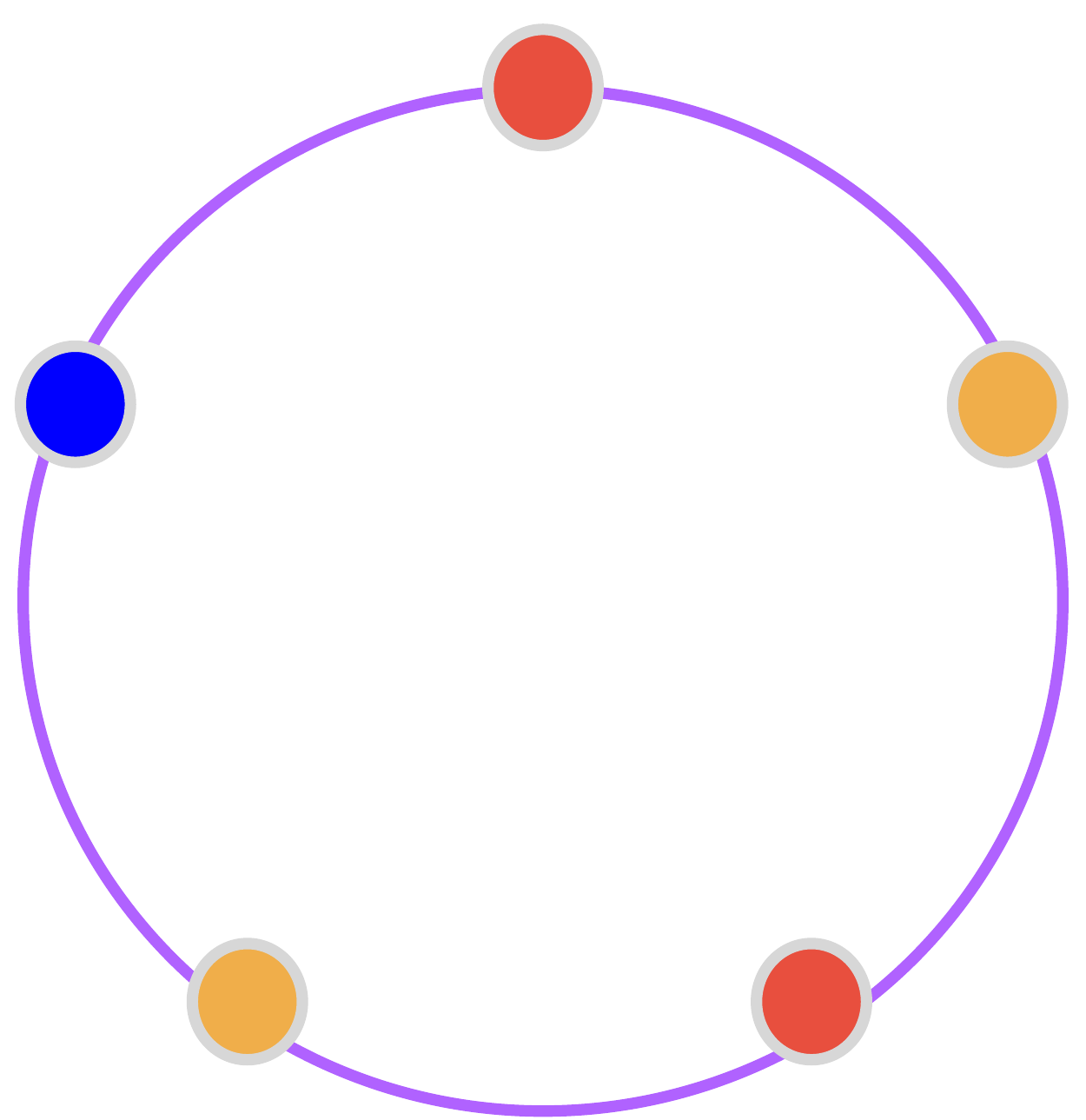} \ \ \ \ \ \ \ \ \ \ \ \ \ \ \ \ \ \ \ \ \ \ \ \ \ \ \
		\includegraphics[scale=0.46]{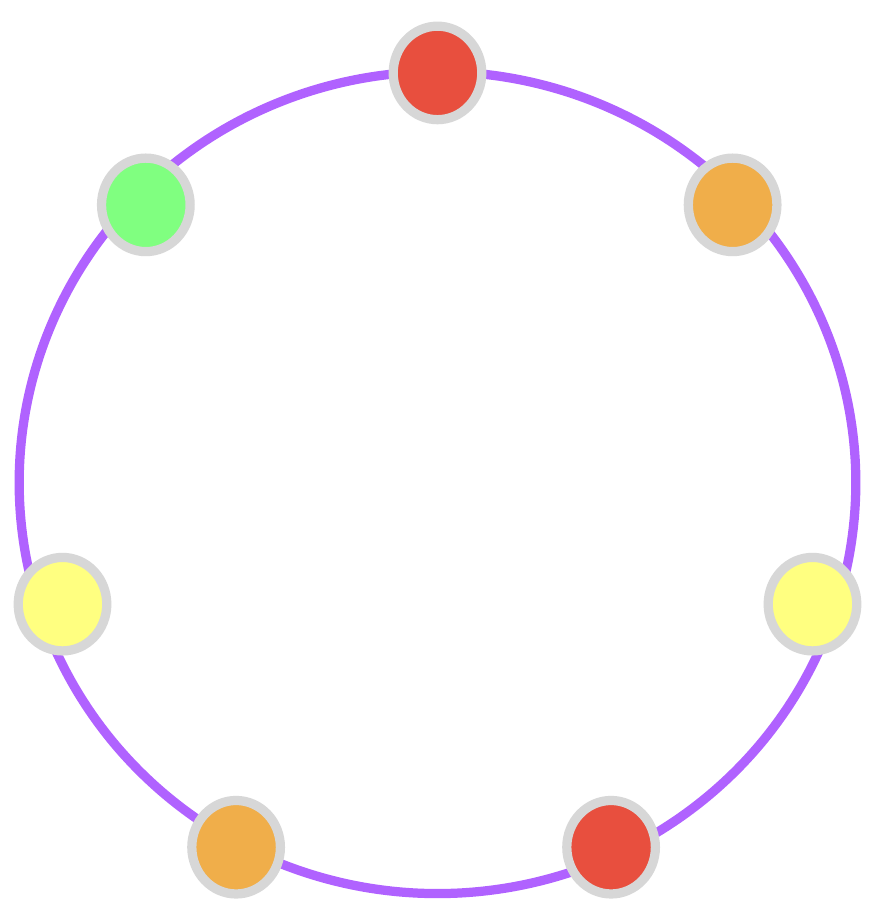}
	\caption{This figure shows clearly why states of the form (\ref{5PME})  and  (\ref{7PME}) are PMEs.  }\label{PMEBell}
\end{figure}

\begin{example} The following state is a PMEs in $(\mathbb{C}^2)^{\otimes 5}$ (see Fig. \ref{PMEBell})
\begin{equation}\label{5PME}
|\Psi\rangle=\frac{1}{2}\sum_{i,j=0}^{1} |i,j,i,j,i\oplus j\rangle.
\end{equation}

\end{example}
\noindent The key point  here is that we have the following equalities
$$\begin{array}{l}
\{(i, j) |i,j=0,1\}=\{(0,0),(0,1),(1,0),(1,1)\},\\[2mm]
\{(j,i\oplus j) |i,j=0,1\}=\{(0,0),(0,1),(1,0),(1,1)\},\\[2mm]
\{( i\oplus j,i) |i,j=0,1\}=\{(0,0),(0,1),(1,0),(1,1)\}.
\end{array}$$
\begin{example} The following state is a PMEs in $(\mathbb{C}^2)^{\otimes 7}$ (see Fig. \ref{PMEBell})
\begin{equation}\label{7PME}
|\Psi\rangle=\frac{1}{2^{3/2}}\sum_{i,j,k=0}^{1} |i,j,k,i,j,k,i\oplus j\oplus k\rangle.
\end{equation}

The key point is   the following equalities
$$
\begin{array}{l}
 \{(i,j,k)\big | \ i,j,k=0,1\}=\{0,1\}^3,\\[2mm]
 \{(j,k,i \oplus j \oplus k)\big | \ i,j,k=0,1\}=\{0,1\}^3,\\[2mm]
 \{(k,i \oplus j \oplus k,i)\big | \ i,j,k=0,1\}=\{0,1\}^3,\\[2mm]
 \{(i \oplus j \oplus k,i,j)\big | \ i,j,k=0,1\}=\{0,1\}^3.
    \end{array}
    $$

\end{example}

\begin{theorem}
Let $N=2n+1$. The  state $|\Phi_o\rangle $ defined in Eq. (\ref{PMEodd})  is a PMEs in $(\mathbb{C}^d)^{\otimes N}$
\begin{equation}\label{PMEodd}
 |\Phi_o\rangle:=\frac{1}{d^{n/2}}\sum_{i_1,\cdots,i_n=0}^{d-1} (\otimes_{k=1}^n|i_k\rangle_k)\otimes(\otimes_{k=1}^n|i_k\rangle_{n+k})\otimes(|\oplus_{k=1}^n i_k\rangle_N).
\end{equation}
\end{theorem}

\noindent\emph{Proof. }
By the symmetry, it is enough to show that
{ $$
\mathcal{S}_1:=\{(i_2,\cdots,i_n,\oplus_{k=1}^ni_k) | i_1,\cdots, i_n=0,\cdots,d-1\}$$
}
 is equal to $\mathcal{S}=\{0,\cdots,d-1\}^n.$ Obviously, we have
{ $$
\mathcal{S}_0:=\{(i_1,\cdots,i_n) | i_1,\cdots, i_n=0,\cdots,d-1\}=\mathcal{S}.$$
}
\noindent We can define a map  $\mathcal{M} $ from $\mathcal{S}_0$ to $\mathcal{S}_1$  which is defined by sending $(i_1,\cdots,i_n)$ to  $(i_2,\cdots,i_n,\oplus_{k=1}^ni_k)$. From the definition of $\mathcal{M} $, one can easily show that if  $$\mathcal{M}[(i_1,\cdots,i_n)] =\mathcal{M}[(i_1',\cdots,i_n')],$$  then $i_2=i_2', \cdots, i_n=i_n'$ and $\oplus_{k=1}^ni_k=\oplus_{k=1}^ni_k'$. Therefore, $i_1=i_1'$. So $\mathcal{M} $ is an injection. Hence the number of elements in $\mathcal{S}_1$ must be larger than  or equal to that of $\mathcal{S}_0$, i.e. $d^n$. With this in mind, we must have $\mathcal{S}_1=\mathcal{S}.$ \qed

\vskip 5pt 

 For any  pure state $|\Psi\rangle$ of bipartite system $\mathcal{H}_A\otimes \mathcal{H}_B:=\mathbb{C}^{d_1}\otimes\mathbb{C}^{d_2}$, it can be written as the form
$$|\Psi\rangle_{A|B}=\sum_{j=1}^k \sqrt{\lambda_j} |e_j\rangle_A|e_j\rangle_B,   \  \sum_{j=1}^k \lambda_j=1,$$
where $\lambda_j>0$ and $\{|e_j\rangle_A\}_{j=1}^k (\{|e_j\rangle_B\}_{j=1}^k)$ are orthonormal states of subsystem $A$ (resp. $B$).  Von Neumann entropy which is defined as $S(|\Psi\rangle_{A|B}):=- \sum_{j=1}^k \lambda_j \log \lambda_j$, is a  quantity that measure the entanglement of $|\Psi\rangle$   (See Ref. \cite{nils}).  The state $|\Psi\rangle$ is entangled if and only if   $S(|\Psi\rangle_{A|B})>0$. Moreover,   $|\Psi\rangle$ is a maximally entangled state if and only if $S(|\Psi\rangle_{A|B})=\log m$ where $m:=\min\{d_1,d_2\}$. In the following, we give a brief  discussion  on the entanglement  of $|\Phi_e\rangle$ and $|\Phi_o\rangle$ in Eq.\eqref{PMEeven} and Eq.\eqref{PMEodd}. 

First, let us consider the state  $|\Phi_e\rangle$. Let  $\{A_1,A_2,\cdots, A_n,A_{n+1},A_{n+2},\cdots, A_{n+n}\}$ denote the $2n$ of the subsystems along the circle. Let $\mathcal{A}$ denote   any  adjacent $n$ subsystems and $\mathcal{B}$ denote all the other subsystems except those in $\mathcal{A}$. As $|\Phi_e\rangle$ is a planar maximally entangled state, $|\Phi_e\rangle_{\mathcal{A}|\mathcal{B}}$ is maximally entangled. Therefore,   we have $S(|\Phi_e\rangle_{\mathcal{A}|\mathcal{B}})=n\log d$.
However, there are some bipartition such that  $|\Phi_e\rangle$ is a product state across that bipartition. For example, set $\mathcal{A}:=\{A_1,A_{n+1}\}$ and $\mathcal{B}:=\{A_i,A_{n+i}\}_{i=2}^n$. Then $|\Phi_e\rangle =|\phi_+\rangle_{A_1,A_{n+1}} \bigotimes (\otimes_{j=2}^n|\phi_+\rangle_{A_j,A_{n+j}})$. Therefore,  $S(|\Phi_e\rangle_{\mathcal{A}|\mathcal{B}})=0$ in such case. Generally, let $\mathcal{A}|\mathcal{B}$ denote a bipartition of the $2n$ subsystems. $A_i$ or $A_{n+i}$ is called a  pairable  partite in   $\mathcal{A}$  ($\mathcal{B}$) if and only if both $A_i$ and $A_{n+i}$ are in $\mathcal{A}$  ($\mathcal{B}$). We denote $\mathcal{A}_p$ ($\mathcal{B}_p$) to be the set of all  pairable  partite in   $\mathcal{A}$ ($\mathcal{B}$).  Moreover, we denote $\mathcal{A}_u:=\mathcal{A}\setminus\mathcal{A}_p$ and  $\mathcal{B}_u:=\mathcal{B}\setminus\mathcal{B}_p$. In the following, we also use the notation $\overline{A}_i:=A_{n+i}$ and $\overline{A}_{n+i}:=A_i$ for $i=1,2,\cdots,n$. Under this notation, we have $A_i\in \mathcal{A}_u$ if and only if $\overline{A}_i\in \mathcal{B}_u$ and $|\mathcal{A}_u|=|\mathcal{B}_u|$ (i.e., $\mathcal{A}_u$ and $\mathcal{B}_u$ have the same number of elements).   Set $s:=|\mathcal{A}_u|$.  Then we have  $S(|\Phi_e\rangle_{\mathcal{A}|\mathcal{B}})=s \log d$. 
In fact,   if $\mathcal{A}_u=\{A_{j_1},\cdots,A_{j_s}\}$, then 
$$|\Phi_e\rangle_{\mathcal{A}|\mathcal{B}}=\frac{1}{d^{s/2}}\sum_{i_{j_1},\cdots,i_{j_s}=0}^{d-1}|\Phi\rangle_{\mathcal{A}_p}|i_{j_1}\rangle_{A_{j_1}}\cdots|i_{j_s}\rangle_{A_{j_s}} \bigotimes |\Phi\rangle_{\mathcal{B}_p} |i_{j_1}\rangle_{\overline{A}_{j_1}}\cdots|i_{j_s}\rangle_{\overline{A}_{j_s}}$$
where $|\Phi\rangle_{\mathcal{A}_p}:=\otimes_{\{A_i,\overline{A}_i\}\subseteq\mathcal{A}_p} {|\phi_+\rangle_{A_i,\overline{A}_i}}$ and $|\Phi\rangle_{\mathcal{B}_p}:=\otimes_{\{A_j,\overline{A}_j\}\subseteq\mathcal{B}_p}{|\phi_+\rangle_{A_j,\overline{A}_j}}$.

Now we consider the entanglement of state  $|\Phi_o\rangle$ across some bipartition of the subsystems. Let   $\{A_1,A_2,\cdots, A_n,A_{n+1},$ $A_{n+2},\cdots, A_{n+n},B_1\}$ denote the $2n+1$ of the subsystems along the circle. Let $\mathcal{A}|\mathcal{B}$ denote a bipartition of the $2n+1$ subsystems. Without loss of generality, we can always assume $B_1\in\mathcal{B}$. Using the same notations $\mathcal{A}_p,\mathcal{B}_p,\mathcal{A}_u$ and $\mathcal{B}_u$ as above, we have $|\mathcal{B}_u|=|\mathcal{A}_u|+1$ as $B_1\in\mathcal{B}_u$. Set $s:=|\mathcal{A}_u|$ and $2t=|\mathcal{A}_p|$.   One finds that (see Appendix \ref{ApenA}) the  Von Neumann entropy $$S(|\Phi_o\rangle_{\mathcal{A}|\mathcal{B}})=[s+\delta(t)]\log d 
$$
where    $\delta(t)=0$ if $t=0$ and $\delta(t)=1$ if $t\geq 1$.  Note that $s\geq 1$ or $t\geq 1$ for any nontrivial bipartition. Therefore, one always has $s+\delta(t)\geq 1$.  As a consquence,  $S(|\Phi_o\rangle_{\mathcal{A}|\mathcal{B}})\geq \log d>0$. So  the state $|\Phi_o\rangle_{\mathcal{A}|\mathcal{B}}$ is an entangled state for any bipartition $\mathcal{A}|\mathcal{B}$, i.e., $|\Phi_o\rangle$ is a genuinely entangled state \cite{Markiewicz13}.
\section{Planar $k$-uniform states in multipartite of  any dimension }\label{fourth}

Now let us turn to a generalization of planar maximally entangled states. First, note that $k$-uniform state  is an important concept that generalizes the maximally entangled state. Here a pure quantum state of $N$ subsystems with local dimension $d$ is called a \emph{$k$-uniform state} if every  reduction  to  $k$ qudits is  maximally mixed. Naturally, we could  generalize the concept planar maximally entangled states to planar $k$-uniform states, i.e. a state whose collection of any $k$   adjacent particles are in a completely mixed states along the circle of parties.

In fact, the existence of planar maximally entangled states for any possible parties $N$ implies existence of planar $k$-uniform states for any possible parties $N$. However, as the high entanglement of the planar maximally entangled states, the circuit for its preparation  might be complex. And a minimal support  $k$-uniform states (the definition of minimal support can be seen in Ref. \cite{Li19})  is more easily to obtain but can also have some potential application  in   quantum secret sharing. We  start with  two simple examples of planar $k$-uniform states.

\begin{figure}[h]
	\centering
	\includegraphics[scale=0.5]{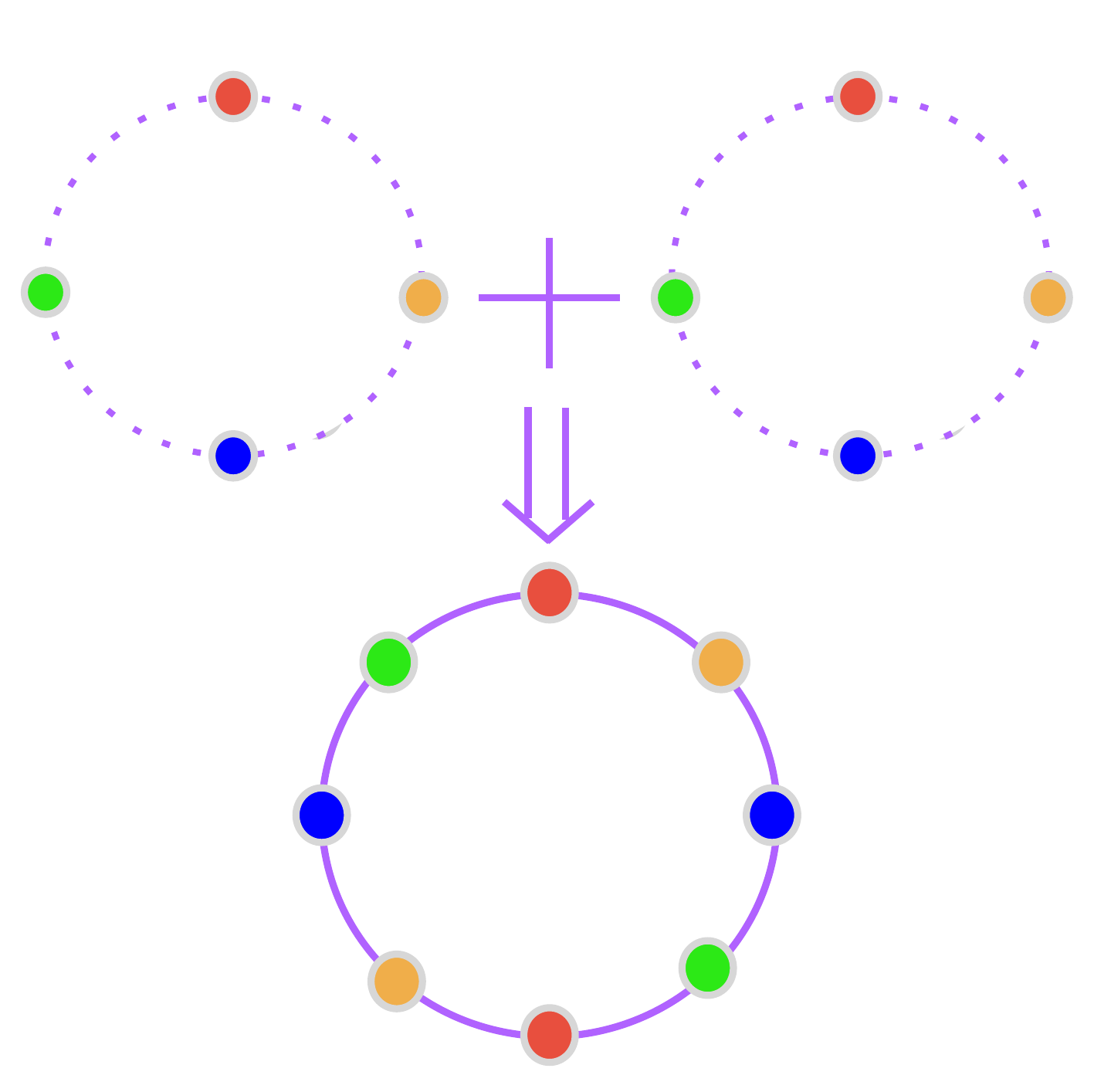}\ \ \ \  \ \ \  \ \ \  \ \ \  \ \ \  \ \ \
\includegraphics[scale=0.5]{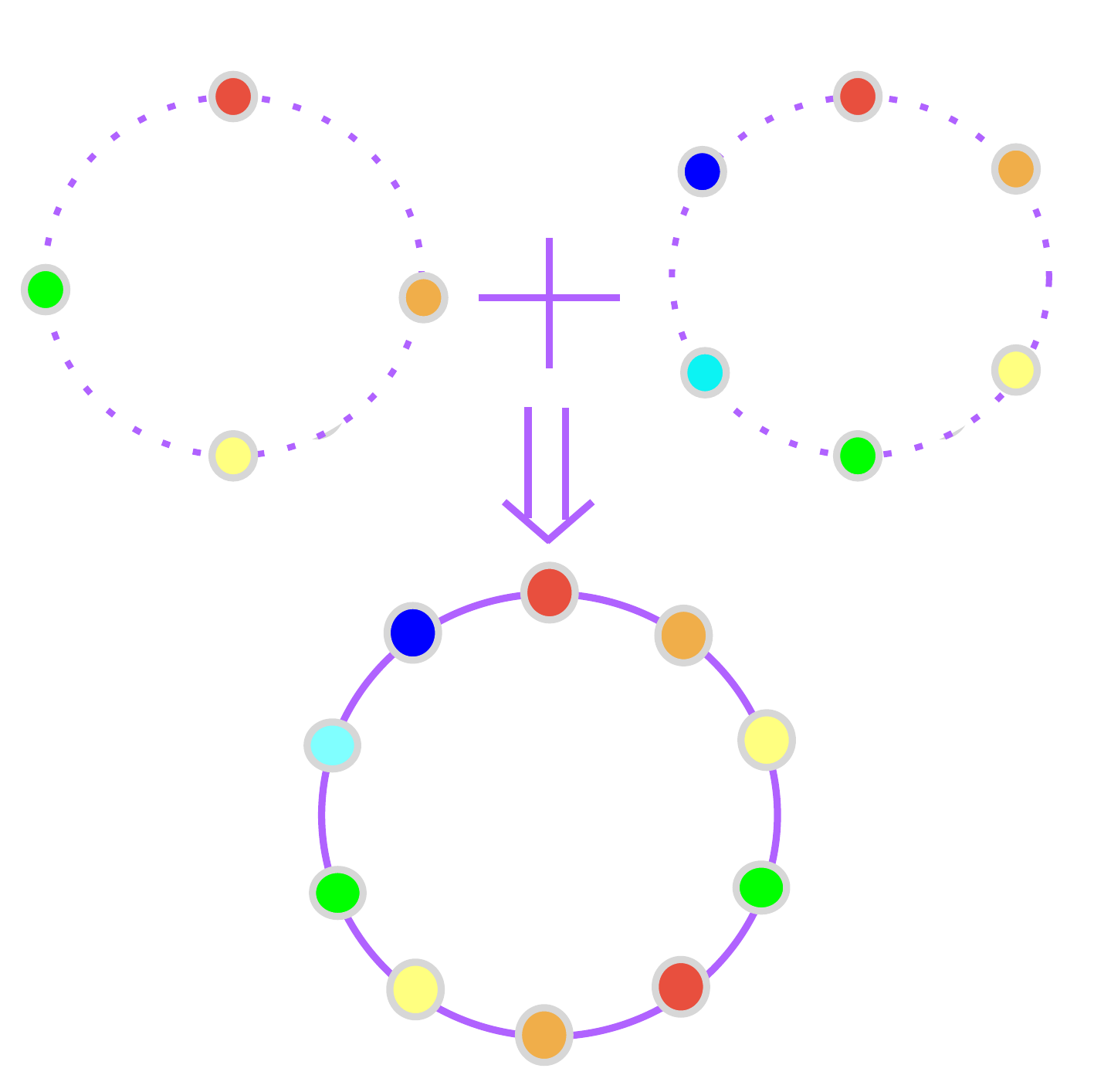}
	\caption{This left figure shows the circle block configuration of the state in Eq. (\ref{3-uniform}).  This right figure shows the circle block configuration of the state in Eq. (\ref{4-uniform}). }\label{PMEBell38}
\end{figure}

\begin{example} The following state is a planar 3-uniform state in $(\mathbb{C}^3)^{\otimes 8}$ (see the left figure of Fig. \ref{PMEBell38})
\begin{equation}\label{3-uniform}
\frac{1}{3^{2}}\sum_{i_1,i_2,i_3=0}^{2} |i_1,i_2,i_3, i_1\oplus i_2\oplus i_3, i_1,i_2,i_3,i_1\oplus i_2\oplus i_3\rangle
\end{equation}
\end{example}

The key point is that the following four sets are all equal to $\{0,1,2\}^3$
$$
\begin{array}{l}
 \{(i_1,i_2,i_3)\big | \ i_1,i_2,i_3=0,1,2\},\\[3mm]
 \{(i_2,i_3, i_1\oplus i_2 \oplus i_3)\big | \ i_1,i_2,i_3 =0,1,2\},\\[3mm]
 \{(i_3,i_1\oplus i_2 \oplus i_3,i_1)\big | \ i_1,i_2,i_3=0,1,2\},\\[3mm]
 \{(i_1\oplus i_2 \oplus i_3,i_1,i_2)\big | \ i_1,i_2,i_3=0,1,2\}.
    \end{array}
    $$

\begin{example}
The following state is a planar 4-uniform state in $(\mathbb{C}^5)^{\otimes 10}$ (see the right figure of Fig. \ref{PMEBell38})
\begin{equation}\label{4-uniform}
\frac{1}{5^{2}}\sum_{i_1,i_2,i_3,i_4=0}^{4} |i_1,i_2,i_3,i_4,   i_1,i_2,i_3,i_4, i_1\oplus i_3,i_2\oplus i_4
\rangle.
\end{equation}
\end{example}

The key point is that the following six sets are all equal to $\{0,1,2,3, 4\}^4$
$$
\begin{array}{l}
 \{(i_1,i_2,i_3,i_4)\big | \ i_1,i_2,i_3,i_4=0,1,2,3, 4\},\\[3mm]
 \{(i_2,i_3,i_4,i_1\oplus i_3)\big | \ i_1,i_2,i_3,i_4=0,1,2,3, 4\},\\[3mm]
 \{(i_3,i_4,i_1\oplus i_3,i_2\oplus i_4)\big | \ i_1,i_2,i_3,i_4=0,1,2,3, 4\},\\[3mm]
 \{(i_4,i_1\oplus i_3,i_2\oplus i_4,i_1)\big | \ i_1,i_2,i_3,i_4=0,1,2,3, 4\},\\[3mm]
 \{(i_1\oplus i_3,i_2\oplus i_4,i_1,i_2)\big | \ i_1,i_2,i_3,i_4=0,1,2,3, 4\},\\[3mm]
 \{( i_2\oplus i_4,i_1,i_2,i_3)\big | \ i_1,i_2,i_3,i_4=0,1,2,3, 4\}.
    \end{array}
$$

For  constructing   $k$-uniform states, Goyeneche \emph{el al.} \cite{Goyeneche14} related them  with a combinatoric natation called orthogonal array.  Now  for  presenting a general method to construct the planar $k$-uniform states, we introduce some parallel notation ``planar orthogonal array".

\vskip 5pt

\noindent{\bf Planar orthogonal array:} An $r \times N$ array $M$ with entries taken
from $\mathbb{Z}_d$ is said to be a \emph{ planar orthogonal array}
with $r$ runs, $N$ factors, $d$ levels, strength $k$, and index $\lambda$ if every  adjacent
$r \times k$ subarray of $M$ contains each $k$-tuple of symbols from $\mathbb{Z}_d$
exactly $\lambda$ times as a row. Here $r$ and $N$ denote the number of
rows and columns of $M$, respectively.  We denote the set of planar orthogonal array of $r$ runs, $N$ factors, $d$ levels, strength $k$ as $\text{POA}(r,N,d,k)$.    A planar orthogonal array $\text{POA}(r,N,d,k)$ is called
 irredundant, if when removing from the array
 any   adjacent $k$ columns all remaining $r$ rows, containing $N-k$ symbols  each, are different. 
 
 Given a   planar orthogonal array $M=(m_{ij})_{i=1,j=1}^{r,N}$ of the form $\text{POA}(r,N,d,k)$, we can always attach it with a pure state $|\Psi_M\rangle$ in $(\mathbb{C}^d)^{\otimes N}$ as 
 $$|\Psi_M\rangle=\frac{1}{\sqrt{r}}\sum_{i=1}^r |m_{i1}\rangle|m_{i2}\rangle\cdots|m_{iN}\rangle.$$
 
One finds that if $M$ is irreducible, then $|\Psi_M\rangle$ is a planar $k$-uniform states.  In the following, we try to
 give a method to construct some irreducible planar orthogonal array of the form $\text{POA}(d^k,N,d,k)$.

\vskip 5pt

\noindent{\bf Circle block of the form $(k,s)$:}  a sequence $(f_1,\cdots, f_{k+s})$ of $k+s$ maps each of which is from $\mathbb{Z}_d^k$ to $\mathbb{Z}_d$ such that any $k$ adjacent maps is a bijection  of $\mathbb{Z}_d^k$. If $f_l(\boldsymbol{i})=i_l$ (where $\boldsymbol{i}=(i_1,\cdots,i_k)\in \mathbb{Z}_d^k$) for $l=1,\cdots, k$, we call it Canonical Circle block. In the following, we use $e_l$ to denote such a map, i.e., the $l$-th coordinate function. If $f,g$ are two maps from $\mathbb{Z}_d^k$ to $\mathbb{Z}_d$, we define $f\oplus g$ to be the  unique map which sends $\boldsymbol{i}$ to $(f(\boldsymbol{i})+g(\boldsymbol{i})) \mod d$.

\vskip 5pt

\noindent\emph{Observation 1. }Given any circle block of the form $(k,s)$, we can construct a $\text{POA}(d^k,k+s,d,k)$ whose rows are consisting of  $ (f_1(\boldsymbol{i}),\cdots, f_{k+s}(\boldsymbol{i})), \boldsymbol{i}\in \mathbb{Z}_d^k$. This is because for any $k$ adjacent coordinates their corresponding map  is  a bijection of $\mathbb{Z}_d^k$. Therefore, any $k$ adjacent coordinates of the defined matrices  consist each $\boldsymbol{i}\in \mathbb{Z}_d^k$ exactly one time.
Moreover, this  $\text{POA}(d^k,k+s,d,k)$ is irredundant if $s\geq k$.
\vskip 5pt

\noindent\emph{Observation 2. }If $(f_1,\cdots, f_{k+s_1})$ and $(g_1,\cdots, g_{k+s_2})$ are  canonical circle blocks of the form $(k,s_1)$ and $(k,s_2)$, then  $(f_1,\cdots, f_{k+s_1},g_1,\cdots,g_{k+s_2} )$ is  also  a canonical circle block, exactly,  of the form $(k,s_1+k+s_2)$.

\vskip 5pt

In order to present the   construction of canonical circle block of general form $(k,s)$, firstly, we  show the constructing process by an example. In the following, we try to construct  a canonical circle block of the form $(11,6)$ of any dimension $d$.
\begin{example}The  following $17$-tuple     is  a canonical circle block of the form $(11,6)$ of any dimension $d$
$$\begin{array}{ll}
(e_1,e_2,\cdots,e_{10},e_{11},e_1\oplus e_{7},e_2\oplus e_{8},e_3\oplus e_{9},\\
e_4\oplus e_{10},e_5\oplus e_{11},e_6\oplus e_{7}\oplus e_{8}\oplus e_{9}\oplus e_{10}\oplus e_{11}).
\end{array}$$
\end{example}

The construction can be separated into three steps as follows.

\vskip 3pt

\noindent \textbf{Step I}. Initializing    a table of $(11+6)\times(11+6)$ as follows (See Table \ref{tab:Step_I})
\begin{enumerate}

   \item[(a)]  The coordinate $(r_i,c_j)$ is filled with $j$ for integers $i,j$ with conditions $1\leq i\leq (11+6)$ , $1\leq j\leq 11$, and $(i-j) \notin\{1,2,3,4,5,6\}.$
   \item[(b)] The coordinate $(r_i,c_{11+j})$  is filled with $j$ for  integers $i,j$ with conditions  $2\leq i\leq 1+6$, $1\leq j\leq 6$, and   $i\leq j$.
\end{enumerate}
One can see that each of the first seven rows are filled with $11$ numbers $\{1,2,\cdots,11\}$.

\vskip 5pt

\begin{table}[h]
\centering
\begin{tabular}{|p{3.5mm}<{\centering}|p{3.3mm}<{\centering}|p{3.3mm}<{\centering}|p{3.3mm}<{\centering}|p{3.3mm}<{\centering}|p{3.3mm}<{\centering}|p{3.3mm}<{\centering}|p{3.3mm}<{\centering}|p{3.3mm}<{\centering}|p{3.3mm}<{\centering}|p{3.3mm}<{\centering}|p{3.3mm}<{\centering}|p{3.3mm}<{\centering}|p{3.3mm}<{\centering}|p{3.3mm}<{\centering}|p{3.3mm}<{\centering}|p{3.3mm}<{\centering}|p{3.3mm}<{\centering}|}
\arrayrulecolor{tabcolor}
\hline
$ V $&$c_1$&$c_2$&$c_3$&$c_4$&$c_5$&$c_6$&$c_7$&$c_8$&$c_9$&$ c_{10}$&$c_{11}$&$c_{12}$&$c_{13}$&$c_{14}$&$c_{15}$&$c_{16}$&$c_{17}$\\ \hline
$r_1$&\textcolor[rgb]{1.00,0.00,0.00}{1}&\textcolor[rgb]{1.00,0.00,0.00}{2}&\textcolor[rgb]{1.00,0.00,0.00}{3}&\textcolor[rgb]{1.00,0.00,0.00}{4}&\textcolor[rgb]{1.00,0.00,0.00}{5}&\textcolor[rgb]{1.00,0.00,0.00}{6}&\textcolor[rgb]{1.00,0.00,0.00}{7}&\textcolor[rgb]{1.00,0.00,0.00}{8}&\textcolor[rgb]{1.00,0.00,0.00}{9}&\textcolor[rgb]{1.00,0.00,0.00}{10}&\textcolor[rgb]{1.00,0.00,0.00}{11}& & & & & &  \\ \hline
$r_2$&&\textcolor[rgb]{1.00,0.00,0.00}{2}&\textcolor[rgb]{1.00,0.00,0.00}{3}&\textcolor[rgb]{1.00,0.00,0.00}{4}&\textcolor[rgb]{1.00,0.00,0.00}{5}&\textcolor[rgb]{1.00,0.00,0.00}{6}&\textcolor[rgb]{1.00,0.00,0.00}{7}&\textcolor[rgb]{1.00,0.00,0.00}{8}&\textcolor[rgb]{1.00,0.00,0.00}{9}&\textcolor[rgb]{1.00,0.00,0.00}{10}&\textcolor[rgb]{1.00,0.00,0.00}{11}& \textcolor[rgb]{1.00,0.00,0.00}{1}& & & & &  \\ \hline
$r_3$&&&\textcolor[rgb]{1.00,0.00,0.00}{3}&\textcolor[rgb]{1.00,0.00,0.00}{4}&\textcolor[rgb]{1.00,0.00,0.00}{5}&\textcolor[rgb]{1.00,0.00,0.00}{6}&\textcolor[rgb]{1.00,0.00,0.00}{7}&\textcolor[rgb]{1.00,0.00,0.00}{8}&\textcolor[rgb]{1.00,0.00,0.00}{9}&\textcolor[rgb]{1.00,0.00,0.00}{10}&\textcolor[rgb]{1.00,0.00,0.00}{11}& \textcolor[rgb]{1.00,0.00,0.00}{1}& \textcolor[rgb]{1.00,0.00,0.00}{2}& & & &  \\ \hline
$r_4$&&&&\textcolor[rgb]{1.00,0.00,0.00}{4}&\textcolor[rgb]{1.00,0.00,0.00}{5}&\textcolor[rgb]{1.00,0.00,0.00}{6}&\textcolor[rgb]{1.00,0.00,0.00}{7}&\textcolor[rgb]{1.00,0.00,0.00}{8}&\textcolor[rgb]{1.00,0.00,0.00}{9}&\textcolor[rgb]{1.00,0.00,0.00}{10}&\textcolor[rgb]{1.00,0.00,0.00}{11}& \textcolor[rgb]{1.00,0.00,0.00}{1}& \textcolor[rgb]{1.00,0.00,0.00}{2}&\textcolor[rgb]{1.00,0.00,0.00}{3}& & &  \\ \hline
$r_5$&&&&&\textcolor[rgb]{1.00,0.00,0.00}{5}&\textcolor[rgb]{1.00,0.00,0.00}{6}&\textcolor[rgb]{1.00,0.00,0.00}{7}&\textcolor[rgb]{1.00,0.00,0.00}{8}&\textcolor[rgb]{1.00,0.00,0.00}{9}&\textcolor[rgb]{1.00,0.00,0.00}{10}&\textcolor[rgb]{1.00,0.00,0.00}{11}& \textcolor[rgb]{1.00,0.00,0.00}{1}& \textcolor[rgb]{1.00,0.00,0.00}{2}&\textcolor[rgb]{1.00,0.00,0.00}{3} & \textcolor[rgb]{1.00,0.00,0.00}{4}& &  \\ \hline
$r_6$&&&&&&\textcolor[rgb]{1.00,0.00,0.00}{6}&\textcolor[rgb]{1.00,0.00,0.00}{7}&\textcolor[rgb]{1.00,0.00,0.00}{8}&\textcolor[rgb]{1.00,0.00,0.00}{9}&\textcolor[rgb]{1.00,0.00,0.00}{10}&\textcolor[rgb]{1.00,0.00,0.00}{11}& \textcolor[rgb]{1.00,0.00,0.00}{1}& \textcolor[rgb]{1.00,0.00,0.00}{2}&\textcolor[rgb]{1.00,0.00,0.00}{3} & \textcolor[rgb]{1.00,0.00,0.00}{4}&\textcolor[rgb]{1.00,0.00,0.00}{5} &  \\ \hline
$r_7$&&&&&&&\textcolor[rgb]{1.00,0.00,0.00}{7}&\textcolor[rgb]{1.00,0.00,0.00}{8}&\textcolor[rgb]{1.00,0.00,0.00}{9}&\textcolor[rgb]{1.00,0.00,0.00}{10}&\textcolor[rgb]{1.00,0.00,0.00}{11}& \textcolor[rgb]{1.00,0.00,0.00}{1}& \textcolor[rgb]{1.00,0.00,0.00}{2}&\textcolor[rgb]{1.00,0.00,0.00}{3} & \textcolor[rgb]{1.00,0.00,0.00}{4}&\textcolor[rgb]{1.00,0.00,0.00}{5} & \textcolor[rgb]{1.00,0.00,0.00}{6}  \\ \hline
$r_8$&\textcolor[rgb]{1.00,0.00,0.00}{1}&&&&&&&\textcolor[rgb]{1.00,0.00,0.00}{8}&\textcolor[rgb]{1.00,0.00,0.00}{9}&\textcolor[rgb]{1.00,0.00,0.00}{10}&\textcolor[rgb]{1.00,0.00,0.00}{11}& & & & & &  \\ \hline
$r_9$&\textcolor[rgb]{1.00,0.00,0.00}{1}&\textcolor[rgb]{1.00,0.00,0.00}{2}&&&&&&&\textcolor[rgb]{1.00,0.00,0.00}{9}&\textcolor[rgb]{1.00,0.00,0.00}{10}&\textcolor[rgb]{1.00,0.00,0.00}{11}& & & & & &  \\ \hline
$r_{10}$&\textcolor[rgb]{1.00,0.00,0.00}{1}&\textcolor[rgb]{1.00,0.00,0.00}{2}&\textcolor[rgb]{1.00,0.00,0.00}{3}&&&&&&&\textcolor[rgb]{1.00,0.00,0.00}{10}&\textcolor[rgb]{1.00,0.00,0.00}{11}& & & & & &  \\ \hline
$r_{11}$&\textcolor[rgb]{1.00,0.00,0.00}{1}&\textcolor[rgb]{1.00,0.00,0.00}{2}&\textcolor[rgb]{1.00,0.00,0.00}{3}&\textcolor[rgb]{1.00,0.00,0.00}{4}&&&&&&&\textcolor[rgb]{1.00,0.00,0.00}{11}& & & & & &  \\ \hline
$r_{12}$&\textcolor[rgb]{1.00,0.00,0.00}{1}&\textcolor[rgb]{1.00,0.00,0.00}{2}&\textcolor[rgb]{1.00,0.00,0.00}{3}&\textcolor[rgb]{1.00,0.00,0.00}{4}&\textcolor[rgb]{1.00,0.00,0.00}{5}&&&&&&& & & & & &  \\ \hline
$r_{13}$&\textcolor[rgb]{1.00,0.00,0.00}{1}&\textcolor[rgb]{1.00,0.00,0.00}{2}&\textcolor[rgb]{1.00,0.00,0.00}{3}&\textcolor[rgb]{1.00,0.00,0.00}{4}&\textcolor[rgb]{1.00,0.00,0.00}{5}&\textcolor[rgb]{1.00,0.00,0.00}{6}&&&&&& & & & & &  \\ \hline
$r_{14}$&\textcolor[rgb]{1.00,0.00,0.00}{1}&\textcolor[rgb]{1.00,0.00,0.00}{2}&\textcolor[rgb]{1.00,0.00,0.00}{3}&\textcolor[rgb]{1.00,0.00,0.00}{4}&\textcolor[rgb]{1.00,0.00,0.00}{5}&\textcolor[rgb]{1.00,0.00,0.00}{6}&\textcolor[rgb]{1.00,0.00,0.00}{7}&&&&& & & & & &  \\ \hline
$r_{15}$&\textcolor[rgb]{1.00,0.00,0.00}{1}&\textcolor[rgb]{1.00,0.00,0.00}{2}&\textcolor[rgb]{1.00,0.00,0.00}{3}&\textcolor[rgb]{1.00,0.00,0.00}{4}&\textcolor[rgb]{1.00,0.00,0.00}{5}&\textcolor[rgb]{1.00,0.00,0.00}{6}&\textcolor[rgb]{1.00,0.00,0.00}{7}&\textcolor[rgb]{1.00,0.00,0.00}{8}&&&& & & & & &  \\ \hline
$r_{16}$&\textcolor[rgb]{1.00,0.00,0.00}{1}&\textcolor[rgb]{1.00,0.00,0.00}{2}&\textcolor[rgb]{1.00,0.00,0.00}{3}&\textcolor[rgb]{1.00,0.00,0.00}{4}&\textcolor[rgb]{1.00,0.00,0.00}{5}&\textcolor[rgb]{1.00,0.00,0.00}{6}&\textcolor[rgb]{1.00,0.00,0.00}{7}&\textcolor[rgb]{1.00,0.00,0.00}{8}&\textcolor[rgb]{1.00,0.00,0.00}{9}&&& & & & & &  \\ \hline
$r_{17}$&\textcolor[rgb]{1.00,0.00,0.00}{1}&\textcolor[rgb]{1.00,0.00,0.00}{2}&\textcolor[rgb]{1.00,0.00,0.00}{3}&\textcolor[rgb]{1.00,0.00,0.00}{4}&\textcolor[rgb]{1.00,0.00,0.00}{5}&\textcolor[rgb]{1.00,0.00,0.00}{6}&\textcolor[rgb]{1.00,0.00,0.00}{7}&\textcolor[rgb]{1.00,0.00,0.00}{8}&\textcolor[rgb]{1.00,0.00,0.00}{9}&\textcolor[rgb]{1.00,0.00,0.00}{10}& & & & & & &  \\ \hline
\end{tabular}
\caption{The initial data of the Step I.}
\label{tab:Step_I}
\end{table}

\noindent\textbf{Step II.} We then recursively define the $j$-th coordinates ($j\geq 12$) of the $(i+1)$-th row  using the result of the $i$-th row when $7\leq i\leq 16$.  Assume that there are exactly $11$ elements named $\{1,2,\cdots, 11\}$ in the $i$-th  row and the non-blank coordinates are $(r_i,c_j)$ with $1\leq j\leq i-7$ and $i\leq j \leq 17$(this is true for $i=7$).  One   finds that the $(i+1)$-th row contains one more element than the $i$-th row  at the left bottom  triangle of the initialized table. In fact, that element is exactly $i-6$ at the coordinate $(r_{i+1},c_{i-6})$.  As  $\{1,2,\cdots, 11\}$ are exactly the elements in the  $i$-th  row, there exists some $x$ such that  $ V (r_i,c_x)=i-6$.  Moreover, as   $ V (r_i,c_j)=j$ for $ 1\leq j \leq i-7$, so $ V (r_i,c_i)\geq i-6$. We will leave the coordinate $(r_{i+1},c_i)$ to be blank.
 \begin{enumerate}
   \item [(1)]If  $ V (r_i,c_i)$ happens to be $i-6$, then we set   $ V (r_{i+1},c_j)= V (r_{i},c_j)$ for $j\geq i+1$. In this setting, we have  $ V (r_{i+1},c_j)= V (r_i,c_j)$ for $1\leq j\leq i-7$ or $i+1\leq j\leq 17$.   In addition, we also have $ V (r_{i+1},c_{i-6})= V (r_{i },c_{i})=i-6.$ Therefore, there are also exactly $11$ elements named $\{1,2,\cdots, 11\}$ in the $(i+1)$-th  row and the non-blank coordinates are $(r_{i+1},c_j)$ with $1\leq j\leq (i+1)-7$ or $i+1\leq j \leq 17$.

   \item [(2)]If the element of the coordinate ($r_i,c_i$) do not equal to $i-6$, i.e., $ V (r_i,c_i)>i-6$,  one may find that  $x\geq i+1$ and $x\in \{12,13,\cdots,17\}$ (otherwise $ V (r_i,c_x)=x=i-6$, hence $i-x=6$ which is contradicted with the condition (a) of step I).  Then we set
$$\ \ \ \ \ \ \  V (r_{i+1},c_j)=\left\{
\begin{array}{ll}
 V (r_{i},c_i),   & j=x\\
 V (r_{i},c_j),&  j\geq i+1 \text{ and } j\neq x.
\end{array}
\right.
$$

In this setting, we have  $ V (r_{i+1},c_j)= V (r_i,c_j)$ for $1\leq j\leq i-7$ or  $i+1\leq j\leq 17$ but $j\neq x$.   In addition, we   have $ V (r_{i+1},c_{i-6})= V (r_{i },c_{x})=i-6$ and $ V (r_{i+1},c_{x})= V (r_{i },c_{i}).$ Therefore, there are also exactly $11$ elements named $\{1,2,\cdots, 11\}$ in the $(i+1)$-th  row and the non-blank coordinates are $(r_{i+1},c_j)$ with $1\leq j\leq (i+1)-7$ or $i+1\leq j \leq 17$. Note that
\begin{equation}\label{turning}
   V (r_{i+1},c_{x})=  V (r_{i },c_{i})>i-6=  V (r_{i},c_{x}).
  \end{equation}
 \end{enumerate}

 Here we present two examples of this recursive definitions as follows.
 \begin{table}[h]
\begin{tabular}{|p{3.3mm}<{\centering}|p{3.3mm}<{\centering}|p{3.3mm}<{\centering}|p{3.3mm}<{\centering}|p{3.3mm}<{\centering}|p{3.3mm}<{\centering}|p{3.3mm}<{\centering}|p{3.3mm}<{\centering}|p{3.3mm}<{\centering}|p{3.3mm}<{\centering}|p{3.3mm}<{\centering}|p{3.3mm}<{\centering}|p{3.3mm}<{\centering}|p{3.3mm}<{\centering}|p{3.3mm}<{\centering}|p{3.3mm}<{\centering}|p{3.3mm}<{\centering}|p{3.3mm}<{\centering}|}
\arrayrulecolor{tabcolor}
\hline
$ V $&$c_1$&$c_2$&$c_3$&$c_4$&$c_5$&$c_6$&$c_7$&$c_8$&$c_9$&$ c_{10}$&$c_{11}$&$c_{12}$&$c_{13}$&$c_{14}$&$c_{15}$&$c_{16}$&$c_{17}$\\ \hline
$r_7$&&&&&&&\multicolumn{1}{>{\columncolor{mypink}}c|} {\textcolor[rgb]{1.00,0.00,0.00}{7}}&\textcolor[rgb]{1.00,0.00,0.00}{8}&\textcolor[rgb]{1.00,0.00,0.00}{9}&\textcolor[rgb]{1.00,0.00,0.00}{10}&\textcolor[rgb]{1.00,0.00,0.00}{11}& \multicolumn{1}{>{\columncolor{mycyan}}c|} {\textcolor[rgb]{1.00,0.00,0.00}{1}}& \textcolor[rgb]{1.00,0.00,0.00}{2}&\textcolor[rgb]{1.00,0.00,0.00}{3} & \textcolor[rgb]{1.00,0.00,0.00}{4}&\textcolor[rgb]{1.00,0.00,0.00}{5} & \textcolor[rgb]{1.00,0.00,0.00}{6}  \\ \hline
$r_8$&   \multicolumn{1}{>{\columncolor{mycyan}}c|} {\textcolor[rgb]{1.00,0.00,0.00}{1}}&&&&&&&\textcolor[rgb]{1.00,0.00,0.00}{8}&\textcolor[rgb]{1.00,0.00,0.00}{9}&\textcolor[rgb]{1.00,0.00,0.00}{10}&\textcolor[rgb]{1.00,0.00,0.00}{11}& \multicolumn{1}{>{\columncolor{mypink}}c|}{\textcolor[rgb]{1.00,0.00,0.00}{7}}& \textcolor[rgb]{1.00,0.00,0.00}{2}&\textcolor[rgb]{1.00,0.00,0.00}{3} & \textcolor[rgb]{1.00,0.00,0.00}{4}&\textcolor[rgb]{1.00,0.00,0.00}{5} & \textcolor[rgb]{1.00,0.00,0.00}{6}  \\ \hline
\end{tabular}
\\[3mm]
\centering
\begin{tabular}{|p{3.3mm}<{\centering}|p{3.3mm}<{\centering}|p{3.3mm}<{\centering}|p{3.3mm}<{\centering}|p{3.3mm}<{\centering}|p{3.3mm}<{\centering}|p{3.3mm}<{\centering}|p{3.3mm}<{\centering}|p{3.3mm}<{\centering}|p{3.3mm}<{\centering}|p{3.3mm}<{\centering}|p{3.3mm}<{\centering}|p{3.3mm}<{\centering}|p{3.3mm}<{\centering}|p{3.3mm}<{\centering}|p{3.3mm}<{\centering}|p{3.3mm}<{\centering}|p{3.3mm}<{\centering}|}
\arrayrulecolor{tabcolor}
\hline
$ V $&$c_1$&$c_2$&$c_3$&$c_4$&$c_5$&$c_6$&$c_7$&$c_8$&$c_9$&$ c_{10}$&$c_{11}$&$c_{12}$&$c_{13}$&$c_{14}$&$c_{15}$&$c_{16}$&$c_{17}$\\ \hline
$r_{13}$&\textcolor[rgb]{1.00,0.00,0.00}{1}&\textcolor[rgb]{1.00,0.00,0.00}{2}&\textcolor[rgb]{1.00,0.00,0.00}{3}&\textcolor[rgb]{1.00,0.00,0.00}{4}&\textcolor[rgb]{1.00,0.00,0.00}{5}&\textcolor[rgb]{1.00,0.00,0.00}{6}&&&&&& & \multicolumn{1}{>{\columncolor{mypink}}c|}{\textcolor[rgb]{1.00,0.00,0.00}{8}}&\textcolor[rgb]{1.00,0.00,0.00}{9} & \textcolor[rgb]{1.00,0.00,0.00}{10}&\textcolor[rgb]{1.00,0.00,0.00}{11} & \multicolumn{1}{>{\columncolor{mycyan}}c|} {\textcolor[rgb]{1.00,0.00,0.00}{7}   }    \\ \hline
$r_{14}$&\textcolor[rgb]{1.00,0.00,0.00}{1}&\textcolor[rgb]{1.00,0.00,0.00}{2}&\textcolor[rgb]{1.00,0.00,0.00}{3}&\textcolor[rgb]{1.00,0.00,0.00}{4}&\textcolor[rgb]{1.00,0.00,0.00}{5}&\textcolor[rgb]{1.00,0.00,0.00}{6}& \multicolumn{1}{>{\columncolor{mycyan}}c|} {\textcolor[rgb]{1.00,0.00,0.00}{7}}&&&&& &  &\textcolor[rgb]{1.00,0.00,0.00}{9} & \textcolor[rgb]{1.00,0.00,0.00}{10}&\textcolor[rgb]{1.00,0.00,0.00}{11} & \multicolumn{1}{>{\columncolor{mypink}}c|}{\textcolor[rgb]{1.00,0.00,0.00}{8} }        \\ \hline

\end{tabular}

\label{tab:number}
\end{table}

Note that  for any fixed integer $j\in [12, 17]$,   the elements in the $j$-th column   are in a nondecreasing  order from top to bottom, i.e. $ V (r_{i+1},c_{j})\geq V (r_{i},c_{j})$ whenever both elements are non-blank.     If $ V (r_{i_2},c_{j})$ is strictly larger than $V (r_{i_1},c_{j})$, we call  $ V (r_{i_1},c_j)$  a predecessor of $ V (r_{i_2},c_j)$ in the $j$-th column and   $ V (r_{i_2},c_j)$   a successor of $ V (r_{i_1},c_j)$ in the $j$-th column.  Denote $\mathcal{P}_j( V (r_{i},c_j))$ (resp. $\mathcal{S}_j( V (r_{i},c_j))$) to be the set of all predecessors (resp. successors) of $ V (r_{i},c_j)$  in the $j$-th column. Fixed $8\leq i+1\leq 17$,     the predecessors   of $ V (r_{i+1},c_j)$ in the $j$-th column ($12\leq j\leq 17$) are contained in the set $\{1,\cdots,i-6\}$. This statement can be followed by the two facts. Fact 1:  there is no  predecessor at all of $ V (r_{7},c_j)$ in the $j$-th column.
 Fact 2: by step II and Eq. (\ref{turning}),  at most one  more predecessor (say $i-6$) would  be generated to one of the element in  $(i+1)$-th row for the recursive definition from $i$-th row to $(i+1)$-th row.

\begin{table}[h]
\centering
\begin{tabular}{|p{3.5mm}<{\centering}|p{3.3mm}<{\centering}|p{3.3mm}<{\centering}|p{3.3mm}<{\centering}|p{3.3mm}<{\centering}|p{3.3mm}<{\centering}|p{3.3mm}<{\centering}|p{3.3mm}<{\centering}|p{3.3mm}<{\centering}|p{3.3mm}<{\centering}|p{3.3mm}<{\centering}|p{3.3mm}<{\centering}|p{3.3mm}<{\centering}|p{3.3mm}<{\centering}|p{3.3mm}<{\centering}|p{3.3mm}<{\centering}|p{3.3mm}<{\centering}|p{3.3mm}<{\centering}|}
\arrayrulecolor{tabcolor}
\hline
$ V $&$c_1$&$c_2$&$c_3$&$c_4$&$c_5$&$c_6$&$c_7$&$c_8$&$c_9$&$ c_{10}$&$c_{11}$&$c_{12}$&$c_{13}$&$c_{14}$&$c_{15}$&$c_{16}$&$c_{17}$\\ \hline
$r_1$&\textcolor[rgb]{1.00,0.00,0.00}{1}&\textcolor[rgb]{1.00,0.00,0.00}{2}&\textcolor[rgb]{1.00,0.00,0.00}{3}&\textcolor[rgb]{1.00,0.00,0.00}{4}&\textcolor[rgb]{1.00,0.00,0.00}{5}&\textcolor[rgb]{1.00,0.00,0.00}{6}&\textcolor[rgb]{1.00,0.00,0.00}{7}&\textcolor[rgb]{1.00,0.00,0.00}{8}&\textcolor[rgb]{1.00,0.00,0.00}{9}&\textcolor[rgb]{1.00,0.00,0.00}{10}&\textcolor[rgb]{1.00,0.00,0.00}{11}& & & & & &  \\ \hline
$r_2$&&\textcolor[rgb]{1.00,0.00,0.00}{2}&\textcolor[rgb]{1.00,0.00,0.00}{3}&\textcolor[rgb]{1.00,0.00,0.00}{4}&\textcolor[rgb]{1.00,0.00,0.00}{5}&\textcolor[rgb]{1.00,0.00,0.00}{6}&\textcolor[rgb]{1.00,0.00,0.00}{7}&\textcolor[rgb]{1.00,0.00,0.00}{8}&\textcolor[rgb]{1.00,0.00,0.00}{9}&\textcolor[rgb]{1.00,0.00,0.00}{10}&\textcolor[rgb]{1.00,0.00,0.00}{11}& \textcolor[rgb]{1.00,0.00,0.00}{1}& & & & &  \\ \hline
$r_3$&&&\textcolor[rgb]{1.00,0.00,0.00}{3}&\textcolor[rgb]{1.00,0.00,0.00}{4}&\textcolor[rgb]{1.00,0.00,0.00}{5}&\textcolor[rgb]{1.00,0.00,0.00}{6}&\textcolor[rgb]{1.00,0.00,0.00}{7}&\textcolor[rgb]{1.00,0.00,0.00}{8}&\textcolor[rgb]{1.00,0.00,0.00}{9}&\textcolor[rgb]{1.00,0.00,0.00}{10}&\textcolor[rgb]{1.00,0.00,0.00}{11}& \textcolor[rgb]{1.00,0.00,0.00}{1}& \textcolor[rgb]{1.00,0.00,0.00}{2}& & & &  \\ \hline
$r_4$&&&&\textcolor[rgb]{1.00,0.00,0.00}{4}&\textcolor[rgb]{1.00,0.00,0.00}{5}&\textcolor[rgb]{1.00,0.00,0.00}{6}&\textcolor[rgb]{1.00,0.00,0.00}{7}&\textcolor[rgb]{1.00,0.00,0.00}{8}&\textcolor[rgb]{1.00,0.00,0.00}{9}&\textcolor[rgb]{1.00,0.00,0.00}{10}&\textcolor[rgb]{1.00,0.00,0.00}{11}& \textcolor[rgb]{1.00,0.00,0.00}{1}& \textcolor[rgb]{1.00,0.00,0.00}{2}&\textcolor[rgb]{1.00,0.00,0.00}{3}& & &  \\ \hline
$r_5$&&&&&\textcolor[rgb]{1.00,0.00,0.00}{5}&\textcolor[rgb]{1.00,0.00,0.00}{6}&\textcolor[rgb]{1.00,0.00,0.00}{7}&\textcolor[rgb]{1.00,0.00,0.00}{8}&\textcolor[rgb]{1.00,0.00,0.00}{9}&\textcolor[rgb]{1.00,0.00,0.00}{10}&\textcolor[rgb]{1.00,0.00,0.00}{11}& \textcolor[rgb]{1.00,0.00,0.00}{1}& \textcolor[rgb]{1.00,0.00,0.00}{2}&\textcolor[rgb]{1.00,0.00,0.00}{3} & \textcolor[rgb]{1.00,0.00,0.00}{4}& &  \\ \hline
$r_6$&&&&&&\textcolor[rgb]{1.00,0.00,0.00}{6}&\textcolor[rgb]{1.00,0.00,0.00}{7}&\textcolor[rgb]{1.00,0.00,0.00}{8}&\textcolor[rgb]{1.00,0.00,0.00}{9}&\textcolor[rgb]{1.00,0.00,0.00}{10}&\textcolor[rgb]{1.00,0.00,0.00}{11}& \textcolor[rgb]{1.00,0.00,0.00}{1}& \textcolor[rgb]{1.00,0.00,0.00}{2}&\textcolor[rgb]{1.00,0.00,0.00}{3} & \textcolor[rgb]{1.00,0.00,0.00}{4}&\textcolor[rgb]{1.00,0.00,0.00}{5} &  \\ \hline
$r_7$&&&&&&&\textcolor[rgb]{1.00,0.00,0.00}{7}&\textcolor[rgb]{1.00,0.00,0.00}{8}&\textcolor[rgb]{1.00,0.00,0.00}{9}&\textcolor[rgb]{1.00,0.00,0.00}{10}&\textcolor[rgb]{1.00,0.00,0.00}{11}& \textcolor[rgb]{1.00,0.00,0.00}{1}& \textcolor[rgb]{1.00,0.00,0.00}{2}&\textcolor[rgb]{1.00,0.00,0.00}{3} & \textcolor[rgb]{1.00,0.00,0.00}{4}&\textcolor[rgb]{1.00,0.00,0.00}{5} & \textcolor[rgb]{1.00,0.00,0.00}{6}  \\ \hline
$r_8$&\textcolor[rgb]{1.00,0.00,0.00}{1}&&&&&&&\textcolor[rgb]{1.00,0.00,0.00}{8}&\textcolor[rgb]{1.00,0.00,0.00}{9}&\textcolor[rgb]{1.00,0.00,0.00}{10}&\textcolor[rgb]{1.00,0.00,0.00}{11}& \textcolor[rgb]{1.00,0.00,0.00}{7}& \textcolor[rgb]{1.00,0.00,0.00}{2}&\textcolor[rgb]{1.00,0.00,0.00}{3} & \textcolor[rgb]{1.00,0.00,0.00}{4}&\textcolor[rgb]{1.00,0.00,0.00}{5} & \textcolor[rgb]{1.00,0.00,0.00}{6}  \\ \hline
$r_9$&\textcolor[rgb]{1.00,0.00,0.00}{1}&\textcolor[rgb]{1.00,0.00,0.00}{2}&&&&&&&\textcolor[rgb]{1.00,0.00,0.00}{9}&\textcolor[rgb]{1.00,0.00,0.00}{10}&\textcolor[rgb]{1.00,0.00,0.00}{11}& \textcolor[rgb]{1.00,0.00,0.00}{7}& \textcolor[rgb]{1.00,0.00,0.00}{8}&\textcolor[rgb]{1.00,0.00,0.00}{3} & \textcolor[rgb]{1.00,0.00,0.00}{4}&\textcolor[rgb]{1.00,0.00,0.00}{5} & \textcolor[rgb]{1.00,0.00,0.00}{6}   \\ \hline
$r_{10}$&\textcolor[rgb]{1.00,0.00,0.00}{1}&\textcolor[rgb]{1.00,0.00,0.00}{2}&\textcolor[rgb]{1.00,0.00,0.00}{3}&&&&&&&\textcolor[rgb]{1.00,0.00,0.00}{10}&\textcolor[rgb]{1.00,0.00,0.00}{11}& \textcolor[rgb]{1.00,0.00,0.00}{7}& \textcolor[rgb]{1.00,0.00,0.00}{8}&\textcolor[rgb]{1.00,0.00,0.00}{9} & \textcolor[rgb]{1.00,0.00,0.00}{4}&\textcolor[rgb]{1.00,0.00,0.00}{5} & \textcolor[rgb]{1.00,0.00,0.00}{6}    \\ \hline
$r_{11}$&\textcolor[rgb]{1.00,0.00,0.00}{1}&\textcolor[rgb]{1.00,0.00,0.00}{2}&\textcolor[rgb]{1.00,0.00,0.00}{3}&\textcolor[rgb]{1.00,0.00,0.00}{4}&&&&&&&\textcolor[rgb]{1.00,0.00,0.00}{11}& \textcolor[rgb]{1.00,0.00,0.00}{7}& \textcolor[rgb]{1.00,0.00,0.00}{8}&\textcolor[rgb]{1.00,0.00,0.00}{9} & \textcolor[rgb]{1.00,0.00,0.00}{10}&\textcolor[rgb]{1.00,0.00,0.00}{5} & \textcolor[rgb]{1.00,0.00,0.00}{6}     \\ \hline
$r_{12}$&\textcolor[rgb]{1.00,0.00,0.00}{1}&\textcolor[rgb]{1.00,0.00,0.00}{2}&\textcolor[rgb]{1.00,0.00,0.00}{3}&\textcolor[rgb]{1.00,0.00,0.00}{4}&\textcolor[rgb]{1.00,0.00,0.00}{5}&&&&&&& \textcolor[rgb]{1.00,0.00,0.00}{7}& \textcolor[rgb]{1.00,0.00,0.00}{8}&\textcolor[rgb]{1.00,0.00,0.00}{9} & \textcolor[rgb]{1.00,0.00,0.00}{10}&\textcolor[rgb]{1.00,0.00,0.00}{11} & \textcolor[rgb]{1.00,0.00,0.00}{6}      \\ \hline
$r_{13}$&\textcolor[rgb]{1.00,0.00,0.00}{1}&\textcolor[rgb]{1.00,0.00,0.00}{2}&\textcolor[rgb]{1.00,0.00,0.00}{3}&\textcolor[rgb]{1.00,0.00,0.00}{4}&\textcolor[rgb]{1.00,0.00,0.00}{5}&\textcolor[rgb]{1.00,0.00,0.00}{6}&&&&&& & \textcolor[rgb]{1.00,0.00,0.00}{8}&\textcolor[rgb]{1.00,0.00,0.00}{9} & \textcolor[rgb]{1.00,0.00,0.00}{10}&\textcolor[rgb]{1.00,0.00,0.00}{11} & \textcolor[rgb]{1.00,0.00,0.00}{7}       \\ \hline
$r_{14}$&\textcolor[rgb]{1.00,0.00,0.00}{1}&\textcolor[rgb]{1.00,0.00,0.00}{2}&\textcolor[rgb]{1.00,0.00,0.00}{3}&\textcolor[rgb]{1.00,0.00,0.00}{4}&\textcolor[rgb]{1.00,0.00,0.00}{5}&\textcolor[rgb]{1.00,0.00,0.00}{6}&\textcolor[rgb]{1.00,0.00,0.00}{7}&&&&& &  &\textcolor[rgb]{1.00,0.00,0.00}{9} & \textcolor[rgb]{1.00,0.00,0.00}{10}&\textcolor[rgb]{1.00,0.00,0.00}{11} & \textcolor[rgb]{1.00,0.00,0.00}{8}         \\ \hline
$r_{15}$&\textcolor[rgb]{1.00,0.00,0.00}{1}&\textcolor[rgb]{1.00,0.00,0.00}{2}&\textcolor[rgb]{1.00,0.00,0.00}{3}&\textcolor[rgb]{1.00,0.00,0.00}{4}&\textcolor[rgb]{1.00,0.00,0.00}{5}&\textcolor[rgb]{1.00,0.00,0.00}{6}&\textcolor[rgb]{1.00,0.00,0.00}{7}&\textcolor[rgb]{1.00,0.00,0.00}{8}& & && &  & & \textcolor[rgb]{1.00,0.00,0.00}{10}&\textcolor[rgb]{1.00,0.00,0.00}{11} & \textcolor[rgb]{1.00,0.00,0.00}{9}          \\ \hline
$r_{16}$&\textcolor[rgb]{1.00,0.00,0.00}{1}&\textcolor[rgb]{1.00,0.00,0.00}{2}&\textcolor[rgb]{1.00,0.00,0.00}{3}&\textcolor[rgb]{1.00,0.00,0.00}{4}&\textcolor[rgb]{1.00,0.00,0.00}{5}&\textcolor[rgb]{1.00,0.00,0.00}{6}&\textcolor[rgb]{1.00,0.00,0.00}{7}&\textcolor[rgb]{1.00,0.00,0.00}{8}&\textcolor[rgb]{1.00,0.00,0.00}{9}& & & &  & & &\textcolor[rgb]{1.00,0.00,0.00}{11} & \textcolor[rgb]{1.00,0.00,0.00}{10}  \\ \hline
$r_{17}$&\textcolor[rgb]{1.00,0.00,0.00}{1}&\textcolor[rgb]{1.00,0.00,0.00}{2}&\textcolor[rgb]{1.00,0.00,0.00}{3}&\textcolor[rgb]{1.00,0.00,0.00}{4}&\textcolor[rgb]{1.00,0.00,0.00}{5}&\textcolor[rgb]{1.00,0.00,0.00}{6}&\textcolor[rgb]{1.00,0.00,0.00}{7}&\textcolor[rgb]{1.00,0.00,0.00}{8}&\textcolor[rgb]{1.00,0.00,0.00}{9}&\textcolor[rgb]{1.00,0.00,0.00}{10}& & &  & & &  & \textcolor[rgb]{1.00,0.00,0.00}{11}\\ \hline
\end{tabular}
\caption{The final data after Step II.}
\label{tab:Step_III}
\end{table}

\vskip 4pt

\noindent\textbf{Step III.} Read out of the non-blank elements for each column. For $j\in \{1,2,\cdots,11+6\}$, set $$\mathcal{I}_j:=\{ V (r_i,c_j) |    V (r_i,c_j)  \text{  is non-blank },  1\leq i\leq 11+6 \}.$$   We   define a $17$-tuple of maps $(f_1,f_2,\cdots,f_{17})$ by  $f_j:=\oplus_{i\in \mathcal{I}_j } e_i$ (here and the following $e_i$ denotes the $i$-th coordinate map from $\mathbb{Z}_d^{11}$ to $\mathbb{Z}_d$).   One finds that $\mathcal{I}_j=\{j\}$ for $1\leq j\leq 11$.  Therefore, $f_j=e_j$ for $ 1\leq j\leq 11$. By the final Table \ref{tab:Step_III},  $(f_1,f_2,\cdots,f_{17})$ can be represented by the coordinate functions as follows
$$\begin{array}{ll}
(e_1,e_2,\cdots,e_{10},e_{11},e_1\oplus e_{7},e_2\oplus e_{8},e_3\oplus e_{9},\\
e_4\oplus e_{10},e_5\oplus e_{11},e_6\oplus e_{7}\oplus e_{8}\oplus e_{9}\oplus e_{10}\oplus e_{11}).
\end{array}$$

\vskip 5pt

We claim that  the $(f_1,f_2,\cdots,f_{17})$  defined above is indeed a canonical circle block of dimensional $d$.
We need to check that any $11$ adjacent maps $(f_i,f_{i+1},\cdots, f_{i+10})$  (where $1\leq i\leq 17$. If the subscript $i+j$ is larger than $17$, we replace the corresponding subscript $i+j$ by $i+j-17$)  is a bijective map from $\mathbb{Z}_d^{11}$ to itself.
It sufficient to show that for any
  $\boldsymbol{i}:=(i_1,i_2,\cdots,i_{11}),$ $\boldsymbol{i}':=(i_1',i_2',\cdots,i_{11}')$ in $ \mathbb{Z}_d^{11},$
 $(f_i(\boldsymbol{i}),f_{i+1}(\boldsymbol{i}),\cdots, f_{i+10}(\boldsymbol{i}))$ equals to $(f_i(\boldsymbol{i}'),f_{i+1}(\boldsymbol{i}'),\cdots, f_{i+10}(\boldsymbol{i}'))$
   implies $ \boldsymbol{i}=\boldsymbol{i}'$.
    Firstly,  we define
$$\begin{array}{rcl}
\mathcal{E}_i:&=&\{l |    V (r_i,c_l)  \text{  is non-blank },  1\leq l\leq 11 \},\\
\mathcal{J}_i:&=&\{l |    V (r_i,c_l)  \text{  is non-blank },  1\leq l\leq 11+6 \}.
\end{array}
$$
Under the above replacement, the set of  subscripts in  $(f_i,f_{i+1},\cdots, f_{i+10})$ is the same with the set $\mathcal{J}_i$.
We separate the argument into three cases.

\begin{enumerate}
  \item [(1)]  $1\leq i\leq 7$.  For $l\in \mathcal{E}_i$, $\mathcal{I}_l=\{l\}$, therefore $f_l=e_l$. Hence $i_l=i_l'$ for such $l$. And for any $l\in \{1,2,\cdots, 11\} \setminus \mathcal{E}_i$, there is exactly one $j$ such that $ V (r_i,c_j)=l$. Moreover, there is no predecessor of $l$ in the $j$-th column and the successors of $l$ in the $j$-th column  are just the set  $\mathcal{I}_j\setminus\{l\}$ whose elements are all greater than or equal to $7$. Hence we have $l\in \mathcal{I}_j$ and $\mathcal{I}_j\setminus\{l\}  \subseteq \mathcal{E}_i$. As $f_j=\oplus_{i\in \mathcal{I}_j}e_i= e_l\oplus(\oplus_{i\in \mathcal{I}_j\setminus\{l\}}e_i)$, so $f_j(\boldsymbol{i})=f_j(\boldsymbol{i}')$ implies that  $i_l=i_l'$. From these, we can conclude that $\boldsymbol{i}=\boldsymbol{i}'$.
  \item [(2)]   $8\leq i\leq 11$. For $l\in \mathcal{E}_i$, $\mathcal{I}_l=\{l\}$, therefore $f_l=e_l$. Hence $i_l=i_l'$ for such $l$. Moreover, one finds that $$\{1,2,\cdots,11\}\setminus \mathcal{E}_i=\{i-6, \cdots, i-1\}.$$
For $l=i-1$, there exists exactly one $j\in[12,17]$ such that $ V (r_i,c_j)=l$.
   We have noted that the predecessors of    $ V (r_i,c_j) $ in the $j$-th column   are in the set $\{1,\cdots, i-7\}\subseteq  \mathcal{E}_i$. As the nondecreasing property of the column,   the  successors of $i-1$ of the $l$ column can only be in $\{i,\cdots,17\}\subseteq \mathcal{E}_i$.   In the expression $f_l$, the $e_l$ is the only one undetermined variable. Hence $i_l=i_l'$. Now for $l=i-2$, there is also exactly one $j\in[12,17]$ such that $ V (r_i,c_j)=l$. We have $\mathcal{P}_j(l)\subseteq\{1,\cdots, i-7\}\subseteq  \mathcal{E}_i$. Moreover, as the nondecreasing property of each column,  $\mathcal{S}_j(l)\subseteq\{i-1,i,\cdots,17\}$. The equality $f_{j}(\boldsymbol{i})=f_{j}(\boldsymbol{i}')$ can be expressed as
    \begin{equation*}
       \ \ \ \ \ i_l \oplus (\oplus_{x\in \mathcal{P}_j(l)\cup \mathcal{S}_j(l)}  i_x ) =i_l' \oplus (\oplus_{x\in \mathcal{P}_j(l)\cup \mathcal{S}_j(l)}  i_x ').
      \end{equation*}
      As we already have $i_x=i_x'$ for all $x\in \mathcal{P}_j(l)\cup\mathcal{S}_j(l)$, hence $i_l=i_l'$.
      This argument is similar for the other undetermined coordinates.  Finally, we would also obtain  $\boldsymbol{i}=\boldsymbol{i}'$.
  \item [(3)]   $11+1\leq i\leq 11+6$. For these cases, $\mathcal{E}_i=\{1,2,\cdots, i-7\}$. So $e_1,\cdots ,e_{i-7}$ are among the list of  $(f_i,f_{i+1},\cdots, f_{i+10})$.  So we always have $i_l=i_l'$ for $1\leq l\leq i-7$. One finds that $$\{1,2,\cdots,11\}\setminus \mathcal{E}_i=\{i-6, \cdots, 11\}.$$ Let $l=11$, there exists some $j\geq 12$ such that $V(r_i,c_j)=l$. We have $\mathcal{P}_j(l)\subseteq \mathcal{E}_i$ and  $\mathcal{S}_j(l)=\emptyset$.  Then  $f_j(\boldsymbol{i})=f_j(\boldsymbol{i}')$ implies that  $i_{11}=i_{11}'$.
      Now for $l=10$, there is also exactly one $j\in[12,17]$ such that $ V (r_i,c_j)=l$. We have $\mathcal{P}_j(l)\subseteq\{1,\cdots, i-7\}\subseteq  \mathcal{E}_i$. Moreover, as the nondecreasing property of each column,  $\mathcal{S}_j(l)\subseteq\{11\}$. The equality $f_{j}(\boldsymbol{i})=f_{j}(\boldsymbol{i}')$ can be expressed as
    \begin{equation*}
       \ \ \ \ \ i_l \oplus (\oplus_{x\in \mathcal{P}_j(l)\cup \mathcal{S}_j(l)}  i_x ) =i_l' \oplus (\oplus_{x\in \mathcal{P}_j(l)\cup \mathcal{S}_j(l)}  i_x ').
      \end{equation*}
      As we already have $i_x=i_x'$ for all $x\in \mathcal{P}_j(l)\cup\mathcal{S}_j(l)$, hence $i_l=i_l'$.
      This argument is similar for the other undetermined coordinates.  Finally, we would also obtain  $\boldsymbol{i}=\boldsymbol{i}'$. \qed
\end{enumerate}


\begin{lemma}\label{canonical_circle}
 Given any positive integers $k,d \geq 2$, for any   $s$ in  $[0,k-1]\cap \mathbb{N}$, there exist some canonical circle block of the form $(k,s)$ with dimensional $d$.
\end{lemma}

See Appendix \ref{app_a} for the proof of this lemma.

\begin{theorem}\label{planar_k-uniform}
 Let $k,d\geq 2$ be     integers. For any integer $N\geq 2k$, there exist some planar $k$-uniform state of minimal support with respect to the circle graph in the $N$ particles system $  (\mathbb{C}^d)^{\otimes N}$.

\end{theorem}

\noindent\emph{ Proof.}  If $k=1$, the well known generalized GHZ state is an $1$-uniform state which has minimal support. So we can assume $k\geq 2$. For any integer $N\geq 2k$, there exists a unique decomposition
$$N=qk+s,  \text{ where } q,s\in  \mathbb{N}, \text{ and } 0\leq s<k.$$
As $N\geq 2k$ and $0\leq s<k$,  so $q\geq 2$.  Then $N=N_1+N_2$ where $N_1=(q-1)k$ and $N_2=k+s.$ By the Lemma \ref{canonical_circle}, there exists canonical circle  block of the form $(k,s)$ with dimensional $d$.  As $(e_1,e_2,\cdots,e_k)$ is always a  canonical circle block of the form $(k,0)$ with dimensional $d$, therefore by the \emph{Observation 2}, after  pasting  $(q-1)$ canonical circle blocks of the form $(k,0)$ together with a  canonical circle blocks of the form $(k,s)$, we arrive at a  canonical circle block  of the form $(k,N)$ with dimensional $d$. By  the \emph{Observation 1}, using this canonical circle block, we can construct a planar orthogonal array $\text{POA}(d^k,N,d,k)$ which is irredundant.  Therefore, there is always   a planar $k$-uniform state in the $N$ parties systems which is minimal support.
\qed

\begin{corollary}
For any positive integers $k$ and $N\geq 2k$, there are at least $\lfloor \frac{N}{2} \rfloor-k +1$  classes  of planar $k$-uniform states which  are    inequivalent under local unitary transformation.
\end{corollary}

\noindent\emph{Proof.}  By Theorem \ref{planar_k-uniform}, we can construct a minimal support  planar $K$-uniform  state $|\Psi_K\rangle $ on $N$ parties whenever $K\leq \lfloor \frac{N}{2} \rfloor$.   Such a state is planar $k$-uniform if and only if $ k\leq K$. Hence $\{|\Psi_K\rangle \ |\  k\leq K\leq  \lfloor \frac{N}{2} \rfloor\} $ is a set of planar $k$-uniform states. A planar $K$-uniform state always has at least $d^{K}$ terms. Therefore, if $K_1<K_2$, a minimal support planar $K_1$-uniform state cannot be a planar $K_2$-uniform state. However, the local unitary transformation preserves the $K$-uniformality. Therefore, $|\Psi_{K_1}\rangle$ and   $|\Psi_{K_2}\rangle$ are equivalent  under local unitary transformation if and only if $K_1=K_2$. \qed

\vskip 5pt

\vskip 5pt

We define $X_d$ and $Z_d$ through their action on this local elements of the
basis states $|j\rangle$ via
$$X_d|j\rangle=|j\oplus1\rangle, Z_d|j\rangle=\omega^j|j\rangle$$
 where  $j\oplus k$  denote the module addition in $\mathbb{Z}_d$ and $\omega$ is a primitive $d$-root of unity, that is,  $\omega=e^{\frac{2\pi \sqrt{-1}}{d}}.$
  We define a local unitary operator
$$U(\vec{a}):=Z_d^{a_1}\otimes \cdots \otimes Z_d^{a_k}\otimes X_d^{a_{k+1}}\otimes\cdots \otimes X_d^{a_{N}}$$
for each $\vec{a}=(a_1,a_2,...,a_N)\in \mathbb{Z}_d^N$. Similar with the the minimal support $k$-uniform basis constructed in Ref. \cite{Li19}, we have the following proposition whose proof is similar to that of Ref. \cite{Li19}.

\begin{proposition}\label{pro1}
Let $|\Psi\rangle$  be a minimal support planar $k$-uniform $N$-qudit state  in $\mathcal{H}=\otimes_{j=1}^N \mathcal{H}_j$ where $\text{dim}_\mathbb{C}\mathcal{H}_j=d$ for each $j$ ($j=1,\cdots,N$).  Under the above notation, the set
{\small$$\mathcal{B}:=\{U(\vec{a})|\Psi\rangle \ \  \big | \ \ \vec{a} \in \mathbb{Z}_d^N\}$$}
form an orthogonal basis  consisting of planar $k$-uniform $N$-qudit states.
\end{proposition}


\section{ Conclusions}\label{fifth}
In this paper, we mainly study some special entangled states for system whose subsystems consisting of  a circle as its underlying topology.
Firstly, we give a construction of  planar maximally entangled states  for the system with  odd number of particles.  This is a supplement to those results in \cite{Doroudiani20}. Then we introduce a more general concept called planar $k$-uniform states.  Inspired by the relation of orthogonal array  with  $k$-uniform state,  we develop a method to construct series of planar orthogonal arrays which are helpful for constructing  planar $k$-uniform states with minimal support.  Similar with the planar maximally entangled states, planar $k$-uniform states are also useful for the quantum  secret sharing in some special quantum setting.

\vskip 5pt

	\noindent{\bf Acknowledgments}\, \,  The author  is very grateful to the reviewers for providing
	us many useful  suggestions which have greatly improved the results of our paper. The author thanks Mao-Sheng Li for helpful discussion on the constructions of canonical circle block. This  work  is supported  by  National  Natural  Science  Foundation  of  China  with Grant No. 11901084, No. 61773119    and
	the Research startup funds of DGUT with Grant No.
	GC300501-103.
	
\appendix

\section{The proof of the Von Neumann entropy for $|\Phi_o\rangle_{\mathcal{A}|\mathcal{B}}$}\label{ApenA}
{\bf Case I: }  $t=0$. If   $\mathcal{A}_u=\{A_{j_1},\cdots,A_{j_s}\}$ and $\mathcal{B}_p=\{A_{i_{j_x}},\overline{A}_{i_{j_x}}\}_{{x=s+1}}^n$, then
$$|\Phi_o\rangle_{\mathcal{A}|\mathcal{B}}=\frac{1}{d^{s/2}}\sum_{i_{j_1},\cdots,i_{j_s}=0}^{d-1}\left( |i_{j_1}\rangle_{A_{j_1}}\cdots|i_{j_s}\rangle_{A_{j_s}} \bigotimes  |e_{i_{j_1},\cdots,i_{j_s}}\rangle_{\mathcal{B}}\right) $$
where $|e_{i_{j_1},\cdots,i_{j_s}}\rangle_{\mathcal{B}}
:=\frac{1}{d^{(n-s)/2}}\sum_{i_{j_{s+1}},\cdots,i_{j_{n}}=0}^{d-1} |i_{j_1}\rangle_{\overline{A}_{j_1}}\cdots|i_{j_s}\rangle_{\overline{A}_{j_s}}(\otimes_{x=s+1}^n|i_{j_x}i_{j_x}\rangle_{ {A}_{j_x},\overline{A}_{j_x}}) |\oplus_{y=1}^n i_{j_y}\rangle_{B_1}.$ One can check that the set  $\{|e_{i_{j_1},\cdots,i_{j_s}}\rangle_{\mathcal{B}}\}_{i_{j_1},\cdots,i_{j_s}=0}^{d-1}$ is an orthonormal set. Therefore,   we have $S(|\Phi_o\rangle_{\mathcal{A}|\mathcal{B}})=s\log d$   for this bipartition.

{\bf Case II:} $t\geq 1.$ If  $\mathcal{A}_u=\{A_{j_1},\cdots,A_{j_s}\}$, $\mathcal{A}_p=\{A_{i_{j_x}},\overline{A}_{i_{j_x}}\}_{{x=s+1}}^{s+t}$ and $\mathcal{B}_p=\{A_{i_{j_y}},\overline{A}_{i_{j_y}}\}_{{y=s+t+1}}^n$,  then we have
\begin{equation}\label{eq:tgeq1}
	|\Phi_o\rangle_{\mathcal{A}|\mathcal{B}}=\frac{1}{d^{(s+1)/2}}\sum_{i_{j_1},\cdots,i_{j_s}=0}^{d-1} \sum_{i=0}^{d-1}\left(|e_{i_{j_1},\cdots,i_{j_s},i}\rangle_{\mathcal{A}}\bigotimes  |e_{i_{j_1},\cdots,i_{j_s},i}\rangle_{\mathcal{B}}\right)
\end{equation}
where $|e_{i_{j_1},\cdots,i_{j_s},i}\rangle_{\mathcal{A}}$ and $|e_{i_{j_1},\cdots,i_{j_s},i}\rangle_{\mathcal{B}}$ are defined as
$$ \begin{array}{ccl}
	|e_{i_{j_1},\cdots,i_{j_s},i}\rangle_{\mathcal{A}}:&=&\frac{1}{d^{(t-1)/2}}\displaystyle \sum_{\oplus_{x=s+1}^{s+t}i_{j_{x}}=i}|i_{j_1}\rangle_{A_{j_1}}\cdots|i_{j_s}\rangle_{A_{j_s}}( \otimes_{x=s+1}^{s+t}|i_{j_x}i_{j_x}\rangle_{ {A}_{j_x},\overline{A}_{j_x}}),\\
	|e_{i_{j_1},\cdots,i_{j_s},i}\rangle_{\mathcal{B}}:&=& \frac{1}{d^{(n-s-t)/2}}\displaystyle\sum_{i_{j_{s+t+1}},\cdots,i_{j_{n}}=0}^{d-1} |i_{j_1}\rangle_{\overline{A}_{j_1}}\cdots|i_{j_s}\rangle_{\overline{A}_{j_s}}(\otimes_{y=s+t+1}^n|i_{j_y}i_{j_y}\rangle_{ {A}_{j_y},\overline{A}_{j_y}}) |(\oplus_{z=1}^s  i_{j_z})\oplus i\oplus (\oplus_{u=s+t+1}^n  i_{j_u})\rangle_{B_1}.
\end{array}
$$  
One can check that the set $\{|e_{i_{j_1},\cdots,i_{j_s},i}\rangle_{\mathcal{A}}\}_{i_{j_1},\cdots,i_{j_s},i=0}^{d-1}$ and $\{|e_{i_{j_1},\cdots,i_{j_s},i}\rangle_{\mathcal{B}}\}_{i_{j_1},\cdots,i_{j_s},i=0}^{d-1}$ are both orthonormal sets. Therefore, by Eq. \eqref{eq:tgeq1}, we have $S(|\Phi_o\rangle_{\mathcal{A}|\mathcal{B}})=(s+1)\log d.$  \qed

\section{The proof of Lemma \ref{canonical_circle}}\label{app_a}\label{ApendixB}

\noindent \textbf{Step I}. Initializing    a table of $(k+s)\times(k+s)$ as follows (See Table \ref{tabgen:Step_I})
\begin{enumerate}

   \item[(a)]  The coordinate $(r_i,c_j)$ is filled with $j$ for integers $i,j$ with conditions $1\leq i\leq (k+s)$ , $1\leq j\leq k$, and $(i-j) \notin\{1,2,\cdots, s\}.$
   \item[(b)] The coordinate $(r_i,c_{k+j})$  is filled with $j$ for  integer $i,j$ with conditions  $2\leq i\leq 1+s$, $1\leq j\leq s$, and   $i\leq j$.
\end{enumerate}
One can see that each of the first seven rows are filled with $k$ numbers $\{1,2,\cdots,k\}$.

\vskip 5pt

\begin{table}[h]
\centering
\begin{tabular}{|p{6.5mm}<{\centering}|p{6.3mm}<{\centering}|p{6.3mm}<{\centering}|p{6.3mm}<{\centering}|p{6.3mm}<{\centering}|p{6.3mm}<{\centering}|p{6.3mm}<{\centering}|p{6.3mm}<{\centering}|p{6.3mm}<{\centering}|p{6.3mm}<{\centering}|p{6.3mm}<{\centering}|}
\arrayrulecolor{tabcolor}
\hline
$V$ & $c_1$&$c_2$&$\cdots$&$c_{s+1}$&$\cdots$&$c_{k-1}$&$c_{k}$&$c_{k+1}$&$\cdots $&$c_{k+s}$\\ \hline
$r_1$& \textcolor[rgb]{1.00,0.00,0.00}{1}& \textcolor[rgb]{1.00,0.00,0.00}{2}&$\cdots$&\textcolor[rgb]{1.00,0.00,0.00}{$s+1$}& $\cdots$ & \textcolor[rgb]{1.00,0.00,0.00}{$k-1$}& \textcolor[rgb]{1.00,0.00,0.00}{$k$}&   & &\\ \hline
$r_2$ & & \textcolor[rgb]{1.00,0.00,0.00}{2}&$\cdots$&\textcolor[rgb]{1.00,0.00,0.00}{$s+1$}& $\cdots$ & \textcolor[rgb]{1.00,0.00,0.00}{$k-1$}& \textcolor[rgb]{1.00,0.00,0.00}{$k$}&\textcolor[rgb]{1.00,0.00,0.00}{1} &   & \\ \hline
$ \vdots$ &   & &$\ddots$& $\vdots$ &$\vdots$& $\vdots$&$\vdots$ &$\vdots$ & $\ddots$&   \\ \hline
$ r_{s+1}$ &   & && \textcolor[rgb]{1.00,0.00,0.00}{$s+1$} & $\cdots$&\textcolor[rgb]{1.00,0.00,0.00}{$k-1$}&\textcolor[rgb]{1.00,0.00,0.00}{$k$}&\textcolor[rgb]{1.00,0.00,0.00}{$1$} & $\cdots$& \textcolor[rgb]{1.00,0.00,0.00}{$s$}  \\ \hline
$ r_{s+2}$ & \textcolor[rgb]{1.00,0.00,0.00}{$1$}  & &&   & $\ddots$&\textcolor[rgb]{1.00,0.00,0.00}{$k-1$}&\textcolor[rgb]{1.00,0.00,0.00}{$k$}&  & & \\ \hline
$ r_{s+3}$ & \textcolor[rgb]{1.00,0.00,0.00}{$1$}  &\textcolor[rgb]{1.00,0.00,0.00}{$2$}  &&   & &\textcolor[rgb]{1.00,0.00,0.00}{$k-1$}&\textcolor[rgb]{1.00,0.00,0.00}{$k$}&  & & \\ \hline
$ \vdots$ & $\vdots$  & $\vdots$&$\ddots$& &  &$\vdots$&$\vdots$&  & & \\ \hline
$ \vdots$ & $\vdots$  & $\vdots$&$\vdots$&$\ddots$ &  &$\vdots$&$\vdots$&  & & \\ \hline
$ \vdots$ & $\vdots$  & $\vdots$&$\vdots$&$\vdots$ &$\ddots$  &$\vdots$&$\vdots$&  & & \\ \hline
$ r_{k+s}$ & \textcolor[rgb]{1.00,0.00,0.00}{$1$}  &  \textcolor[rgb]{1.00,0.00,0.00}{$2$}&$\cdots$&  \textcolor[rgb]{1.00,0.00,0.00}{$s+1$}& $\cdots$&\textcolor[rgb]{1.00,0.00,0.00}{$k-1$}& &  & & \\ \hline

\end{tabular}
\caption{The initial data of the Step I of the proof of Lemma \ref{canonical_circle}.}
\label{tabgen:Step_I}
\end{table}

\noindent\textbf{Step II.} We then recursively define the $j$-th coordinates ($j\geq k+1$) of the $(i+1)$-th row  using the result of the $i$-th row when $s+1\leq i\leq k+s-1$.  Assume that there are exactly $k$ elements named $\{1,2,\cdots, k\}$ in the $i$-th  row and the non-blank coordinates are $(r_i,c_j)$ with $1\leq j\leq i-(s+1)$ and $i\leq j \leq k+s$(this is true for $i=s+1$).  One   finds that the $(i+1)$-th row contains one more element than the $i$-th row  at the left bottom  triangle of the initialized table. In fact, that element is exactly $i-s$ at the coordinate $(r_{i+1},c_{i-s})$.  As  $\{1,2,\cdots, k\}$ are exactly the elements in the  $i$-th  row, there exists some $x$ such that  $ V (r_i,c_x)=i-s$.  Moreover, as   $ V (r_i,c_j)=j$ for $ 1\leq j \leq i-(s+1)$, so $ V (r_i,c_i)\geq i-s$. We will leave the coordinate $(r_{i+1},c_i)$ to be blank.
 \begin{enumerate}
   \item [(1)]If  $ V (r_i,c_i)$ happens to be $i-s$, then we set   $ V (r_{i+1},c_j)= V (r_{i},c_j)$ for $j\geq i+1$. In this setting, we have  $ V (r_{i+1},c_j)= V (r_i,c_j)$ for $1\leq j\leq i-(s+1)$ or $i+1\leq j\leq k+s$.   In addition, we also have $ V (r_{i+1},c_{i-s})= V (r_{i },c_{i})=i-s.$ Therefore, there are also exactly $k$ elements named $\{1,2,\cdots, k\}$ in the $(i+1)$-th  row and the non-blank coordinates are $(r_{i+1},c_j)$ with $1\leq j\leq (i+1)-(s+1)$ or $i+1\leq j \leq k+s$.

   \item [(2)]If the element of the coordinate ($r_i,c_i$) do not equal to $i-s$, i.e., $ V (r_i,c_i)>i-s$,  one may find that  $x\geq i+1$ and $x\in \{k+1,k+2,\cdots,k+s\}$ (otherwise $ V (r_i,c_x)=x=i-s$, hence $i-x=s$ which is contradicted with the condition (a) of step I).  Then we set
$$\ \ \ \ \ \ \  V (r_{i+1},c_j)=\left\{
\begin{array}{ll}
 V (r_{i},c_i),   & j=x\\
 V (r_{i},c_j),&  j\geq i+1 \text{ and } j\neq x.
\end{array}
\right.
$$

In this setting, we have  $ V (r_{i+1},c_j)= V (r_i,c_j)$ for $1\leq j\leq i-(s+1)$ or  $i+1\leq j\leq k+s$ but $j\neq x$.   In addition, we   have $ V (r_{i+1},c_{i-s})= V (r_{i },c_{x})=i-s$ and $ V (r_{i+1},c_{x})= V (r_{i },c_{i}).$ Therefore, there are also exactly $k$ elements named $\{1,2,\cdots, k\}$ in the $(i+1)$-th  row and the non-blank coordinates are $(r_{i+1},c_j)$ with $1\leq j\leq (i+1)-(s+1)$ or $i+1\leq j \leq k+s$. Note that
\begin{equation}\label{turningA}
   V (r_{i+1},c_{x})=  V (r_{i },c_{i})>i-s=  V (r_{i},c_{x}).
  \end{equation}
 \end{enumerate}

Note that  for any fixed integer $j\in [k+1, k+s]$,   the elements in the $j$-th column   are in a nondecreasing  order from top to bottom, i.e. $ V (r_{i+1},c_{j})\geq V (r_{i},c_{j})$ whenever both elements are non-blank.     If $ V (r_{i_2},c_{j})$ is strictly larger than $V (r_{i_1},c_{j})$, we call  $ V (r_{i_1},c_j)$  a predecessor of $ V (r_{i_2},c_j)$ in the $j$-th column and   $ V (r_{i_2},c_j)$   a successor of $ V (r_{i_1},c_j)$ in the $j$-th column.  Denote $\mathcal{P}_j( V (r_{i},c_j))$ (resp. $\mathcal{S}_j( V (r_{i},c_j))$) to be the set of all predecessors (resp. successors) of $ V (r_{i},c_j)$  in the $j$-th column. Fixed $s+2\leq i+1\leq k+s$,     the predecessors   of $ V (r_{i+1},c_j)$ in the $j$-th column ($k+1\leq j\leq k+s$) are contained in the set $\{1,\cdots,i-s\}$. This statement can be followed by the two facts. Fact 1:  there is no  predecessor at all of $ V (r_{s+1},c_j)$ in the $j$-th column.
 Fact 2: by step II and Eq. (\ref{turningA}),  at most one  more predecessor (say $i-s$) would  be generated to one of the element in  $(i+1)$-th row for the recursive definition from $i$-th row to $(i+1)$-th row.

\vskip 4pt

\noindent\textbf{Step III.} Read out of the non-blank elements for each column. For $j\in \{1,2,\cdots,k+s\}$, set $$\mathcal{I}_j:=\{ V (r_i,c_j) |    V (r_i,c_j)  \text{  is non-blank },  1\leq i\leq k+s \}.$$   We   define a $(k+s)$-tuple of maps $(f_1,f_2,\cdots,f_{k+s})$ by  $f_j:=\oplus_{i\in \mathcal{I}_j } e_i$ (here and the following $e_i$ denotes the $i$-th coordinate map from $\mathbb{Z}_d^{k}$ to $\mathbb{Z}_d$).   One finds that $\mathcal{I}_j=\{j\}$ for $1\leq j\leq k$.  Therefore, $f_j=e_j$ for $ 1\leq j\leq k$.

\vskip 5pt

We claim that  the $(f_1,f_2,\cdots,f_{k+s})$  defined above is indeed a canonical circle block of dimensional $d$.
We need to check that any $k$ adjacent maps $(f_i,f_{i+1},\cdots, f_{i+k-1})$  (where $1\leq i\leq k+s$. If the subscript $i+j$ is larger than $k+s$, we replace the corresponding subscript $i+j$ by $i+j-(k+s)$)  is a bijective map from $\mathbb{Z}_d^{k}$ to itself.
It sufficient to show that for any
  $\boldsymbol{i}:=(i_1,i_2,\cdots,i_{k}),$ $\boldsymbol{i}':=(i_1',i_2',\cdots,i_{k}')$ in $ \mathbb{Z}_d^{k},$
 $(f_i(\boldsymbol{i}),f_{i+1}(\boldsymbol{i}),\cdots, f_{i+k-1}(\boldsymbol{i}))$ equals to $(f_i(\boldsymbol{i}'),f_{i+1}(\boldsymbol{i}'),\cdots, f_{i+k-1}(\boldsymbol{i}'))$
   implies $ \boldsymbol{i}=\boldsymbol{i}'$.
    Firstly,  we define
$$\begin{array}{rcl}
\mathcal{E}_i:&=&\{l |    V (r_i,c_l)  \text{  is non-blank },  1\leq l\leq k \},\\
\mathcal{J}_i:&=&\{l |    V (r_i,c_l)  \text{  is non-blank },  1\leq l\leq k+s \}.
\end{array}
$$
Under the above replacement, the set of  subscripts in  $(f_i,f_{i+1},\cdots, f_{i+k})$ is the same with the set $\mathcal{J}_i$.
We separate the argument into three cases.

\begin{enumerate}
  \item [(1)]  $1\leq i\leq s+1$.  For $l\in \mathcal{E}_i$, $\mathcal{I}_l=\{l\}$, therefore $f_l=e_l$. Hence $i_l=i_l'$ for such $l$. And for any $l\in \{1,2,\cdots, k\} \setminus \mathcal{E}_i$, there is exactly one $j$ such that $ V (r_i,c_j)=l$. Moreover, there is no predecessor of $l$ in the $j$-th column and the successors of $l$ in the $j$-th column  are just the set  $\mathcal{I}_j\setminus\{l\}$ whose elements are all greater than or equal to $s+1$. Hence we have $l\in \mathcal{I}_j$ and $\mathcal{I}_j\setminus\{l\}  \subseteq \mathcal{E}_i$. As $f_j=\oplus_{i\in \mathcal{I}_j}e_i= e_l\oplus(\oplus_{i\in \mathcal{I}_j\setminus\{l\}}e_i)$, so $f_j(\boldsymbol{i})=f_j(\boldsymbol{i}')$ implies that  $i_l=i_l'$. From these, we can conclude that $\boldsymbol{i}=\boldsymbol{i}'$.
  \item [(2)]   $s+2\leq i\leq k$. For $l\in \mathcal{E}_i$, $\mathcal{I}_l=\{l\}$, therefore $f_l=e_l$. Hence $i_l=i_l'$ for such $l$. Moreover, one finds that $$\{1,2,\cdots,k\}\setminus \mathcal{E}_i=\{i-s, \cdots, i-1\}.$$
For $l=i-1$, there exists exactly one $j\in[k+1,k+s]$ such that $ V (r_i,c_j)=l$.
   We have noted that the predecessors of    $ V (r_i,c_j) $ in the $j$-th column   are in the set $\{1,\cdots, i-(s+1)\}\subseteq  \mathcal{E}_i$. As the nondecreasing property of the column,   the  successors of $i-1$ of the $l$ column can only be in $\{i,\cdots,k+s\}\subseteq \mathcal{E}_i$.   In the expression $f_l$, the $e_l$ is the only one undetermined variable. Hence $i_l=i_l'$. Now for $l=i-2$, there is also exactly one $j\in[k+1,k+s]$ such that $ V (r_i,c_j)=l$. We have $\mathcal{P}_j(l)\subseteq\{1,\cdots, i-(s+1)\}\subseteq  \mathcal{E}_i$. Moreover, as the nondecreasing property of each column,  $\mathcal{S}_j(l)\subseteq\{i-1,i,\cdots,k+s\}$. The equality $f_{j}(\boldsymbol{i})=f_{j}(\boldsymbol{i}')$ can be expressed as
    \begin{equation*}
       \ \ \ \ \ i_l \oplus (\oplus_{x\in \mathcal{P}_j(l)\cup \mathcal{S}_j(l)}  i_x ) =i_l' \oplus (\oplus_{x\in \mathcal{P}_j(l)\cup \mathcal{S}_j(l)}  i_x ').
      \end{equation*}
      As we already have $i_x=i_x'$ for all $x\in \mathcal{P}_j(l)\cup\mathcal{S}_j(l)$, hence $i_l=i_l'$.
      This argument is similar for the other undetermined coordinates.  Finally, we would also obtain  $\boldsymbol{i}=\boldsymbol{i}'$.
  \item [(3)]   $k+1\leq i\leq k+s$. For these cases, $\mathcal{E}_i=\{1,2,\cdots, i-(s+1)\}$. So $e_1,\cdots ,e_{i-(s+1)}$ are among the list of  $(f_i,f_{i+1},\cdots, f_{i+k-1})$.  So we always have $i_l=i_l'$ for $1\leq l\leq i-(s+1)$. One finds that $$\{1,2,\cdots,k\}\setminus \mathcal{E}_i=\{i-s, \cdots, k\}.$$ Let $l=k$, there exists some $j\geq k+1$ such that $V(r_i,c_j)=l$. We have $\mathcal{P}_j(l)\subseteq \mathcal{E}_i$ and  $\mathcal{S}_j(l)=\emptyset$.  Then  $f_j(\boldsymbol{i})=f_j(\boldsymbol{i}')$ implies that  $i_{k}=i_{k}'$.
      For $l=k-1$, there is also exactly one $j\in[k+1,k+s]$ such that $ V (r_i,c_j)=l$. We have $\mathcal{P}_j(l) \subseteq  \mathcal{E}_i$. Moreover, as the nondecreasing property of each column,  $\mathcal{S}_j(l)\subseteq\{k\}$. The equality $f_{j}(\boldsymbol{i})=f_{j}(\boldsymbol{i}')$ can be expressed as
    \begin{equation*}
       \ \ \ \ \ i_l \oplus (\oplus_{x\in \mathcal{P}_j(l)\cup \mathcal{S}_j(l)}  i_x ) =i_l' \oplus (\oplus_{x\in \mathcal{P}_j(l)\cup \mathcal{S}_j(l)}  i_x ').
      \end{equation*}
      As we already have $i_x=i_x'$ for all $x\in \mathcal{P}_j(l)\cup\mathcal{S}_j(l)$, hence $i_l=i_l'$.
      This argument is similar for the other undetermined coordinates.  Finally, we would also obtain  $\boldsymbol{i}=\boldsymbol{i}'$. \qed
\end{enumerate}


\end{document}